\documentclass[a4paper,12pt]{article}
\pdfoutput=1

\usepackage{amsmath}
\usepackage{amssymb}
\usepackage{mathrsfs}
\usepackage{bbm}
\usepackage{graphicx,subfigure}
\usepackage[numbers,sort&compress]{natbib}

\usepackage{color}
\usepackage{ulem}

\numberwithin{equation}{section}

\newlength{\dinwidth}
\newlength{\dinmargin}
\begin{document}

\title{\vspace{-2.2cm}
\bf Exclusive radiative B-meson decays within minimal flavour-violating 2HDMs}
\bigskip
\author{Xin-Qiang Li$^{1,2,3}$\footnote{xqli@itp.ac.cn}, Ya-Dong Yang$^{1,2}$\footnote{yangyd@iopp.ccnu.edu.cn} and Xing-Bo Yuan$^{1,2}$\footnote{xbyuan@iopp.ccnu.edu.cn}\\
{$^1$\small Institute of Particle Physics, Central China Normal University, Wuhan, Hubei 430079, China}\\[-0.2cm]
{$^2$\small Key Laboratory of Quark \& Lepton Physics, Ministry of Education, China}\\[-0.2cm]
{$^3$\small State Key Laboratory of Theoretical Physics, Institute of Theoretical Physics,}\\[-0.2cm]
{    \small Chinese Academy of Sciences, Beijing 100190, China}}

\date{}
\maketitle
\bigskip\bigskip
\maketitle
\vspace{-1.2cm}

\begin{abstract}
{\noindent}In the ``Higgs basis" for a generic 2HDM, only one doublet gets a nonzero vacuum expectation value and, under the criterion of minimal flavour violation, the other one is fixed to be either colour-singlet or colour-octet, referred to, respectively, as the type-III and type-C models. Both of them can naturally avoid large FCNC transitions and provide very interesting phenomena in some low-energy processes. In this paper, we study their effects on exclusive radiative B-meson decays due to the exchange of colourless or coloured charged Higgs. It is found that, while constraints from the branching ratios are slightly weaker than the ones from the inclusive $B\to X_{s} \gamma$ decay, the isospin asymmetries in exclusive decays provide very complementary bounds on the model parameters. As the two models predict similar corrections to the dipole coefficient $C_7^{\rm eff}$, but similar magnitudes with with opposite signs to $C_8^{\rm eff}$, the branching ratios cannot discriminate the two models, and we have to resort to the direct CP and isospin asymmetries of $b\to s$ processes, which are more sensitive to $C_8^{\rm eff}$. Due to the CKM factors $|\lambda_u^{(d)}|\sim|\lambda_t^{(d)}|$, the terms proportional to $\lambda_u^{(d)}$ make the observables of $b\to d$ processes exhibit a different dependence on the possible new physics phase. In addition, correlations between the various observables in the exclusive $B\to V\gamma$ and the inclusive $B\to X_{s,d}\gamma$ decays are investigated, which could provide further insights into the models with more precise experimental measurements and theoretical predictions for these decays.
\end{abstract}

\noindent{{\bf PACS numbers:} 13.20.He, 12.60.Fr, 14.80.Fd}

\newpage

\section{Introduction}
\label{sec:intro}

One of the main goals of the Large Hadron Collider~(LHC) is to explore the mechanism of electroweak symmetry breaking~(EWSB). In the Standard Model~(SM), it is realized via the Higgs mechanism implemented only by one scalar doublet, and the predicted Higgs-boson mass is consistent with the new particle announced by the ATLAS~\cite{Higgs:ATLAS} and CMS~\cite{Higgs:CMS} experiments at LHC. Moreover, its properties measured so far~\cite{Aad:2013wqa,Chatrchyan:2013lba,CMS:yva,Aaltonen:2013kxa} also comply with the ones predicted within the SM. If this boson, with more precise data accumulated, is confirmed to be truly SM-like, a natural question to address is then whether it corresponds to the unique Higgs boson predicted by the SM, or it is just the first signal of a much richer scenario of EWSB.

In fact, the EWSB is not necessarily induced by a single scalar doublet. Meanwhile, the SM by itself is not expected to be a complete description of nature. The simplest extension compatible with the gauge invariance is the so-called two-Higgs-doublet Model~(2HDM)~\cite{Lee}, which is identical to the SM except for one extra scalar doublet. The 2HDM is very interesting on its own as a potential theory of nature, since the extended scalar sector allows for CP violation beyond what is provided by the Cabibbo-Kobayashi-Maskawa~(CKM) mechanism~\cite{Cabibbo:1963yz,Kobayashi:1973fv} in the SM. It is also useful to gain further insights into the scalar sector of supersymmetry and other models that contain similar scalar contents.

Within the SM, the flavour-changing neutral current~(FCNC) interaction is forbidden at tree level and, due to the Glashow-Iliopoulos-Maiani~(GIM) mechanism~\cite{GIM}, is highly suppressed at loop level. In a generic 2HDM, however, the scalar-mediated FCNC transitions are not protected by the GIM mechanism, and will appear unless the off-diagonal couplings of Higgs bosons to quarks are absent or sufficiently small. Accordingly, one big problem 2HDM has to face is how to avoid the stringent experimental constraints on FCNC processes. To address this issue, two different hypotheses, natural flavour conservation~(NFC)~\cite{NFC} and minimal flavour violation~(MFV)~\cite{MFV:1,MFV:2,MFV:3,Buras:2010mh,Cervero:2012cx}, have been proposed. In the NFC hypothesis, by requiring the Yukawa couplings to up and down quarks for all the Higgs fields be diagonal in the basis where the quark mass matrices are diagonal, one can naturally eliminate the tree-level FCNC interactions. Explicitly, this can be enforced via a discrete $Z_2$ symmetry acting differently on the two scalar doublets; depending on the $Z_2$ charge assignments on the scalar doublets and fermions, there are four types of 2HDM~(type-I, II, X and Y) under the NFC hypothesis~\cite{Branco:2011iw,Gunion:1989we}.

In the MFV hypothesis, although being allowed even at tree level, all the flavour-violating interactions, including those mediated by the electrically neutral scalars, are controlled by the CKM matrix~\cite{Cabibbo:1963yz,Kobayashi:1973fv}, as happens in the SM. Explicitly, this can be implemented by requiring all the Higgs Yukawa couplings be composed of the pair of the SM ones $Y^{U}$ and $Y^{D}$. As pointed out in ref.~\cite{Wise}, there are two classes of 2HDM satisfying the MFV hypothesis. For convenience of discussion, we introduce the so-called ``Higgs basis", in which only one doublet gets a nonzero vacuum expectation value~(VEV) and behaves the same as the SM one~\cite{Davidson:2005cw}. In this basis, under the MFV hypothesis, the allowed $SU(3)_C \otimes SU(2)_L \otimes U(1)_Y$ representation of the second doublet is fixed to be either $(\bold 1,\bold 2)_{1/2}$ or $(\bold 8, \bold 2)_{1/2}$; namely, the second doublet can be either colour-singlet or colour-octet~\cite{Wise}, referred to, respectively, as the type-III\footnote{It should be noted that different notations of the type-III 2HDM exist in the literature. The type-III model introduced in this paper denotes the 2HDM under MFV hypothesis with the second Higgs doublet color-singlet, which is defined in detail in section~2. This terminology is, however, usually used for general 2HDMs unconstrained by a $Z_2$ symmetry with the second Higgs doublet colorless, where FCNC is controlled with a particular Yukawa texture~\cite{DiazCruz:2009ek,Mahmoudi:2009zx} or for a decoupling limit of MSSM~\cite{Crivellin:2010er,Crivellin:2012ye}.} and type-C models~\cite{B2Xg:2HDM:0}. Examples of the former include the aligned 2HDM~(A2HDM)~\cite{A2HDM} and the four types of 2HDM reviewed in ref.~\cite{Branco:2011iw}. The scalar spectrum of the latter contains, besides a CP-even and colour-singlet Higgs boson~(the usual SM one), three colour-octet particles, one CP-even, one CP-odd and one electrically charged, providing many interesting phenomena in collider physics~\cite{Wise,Gresham:2007ri,Dobrescu:2007yp,Gerbush:2007fe,Burgess:2009wm,Idilbi:2010rs,Cao:2013wqa}.

Although the scalar-mediated flavour-violating interactions are protected by the MFV hypothesis, these two models still present very interesting phenomena in some low-energy processes, especially due to the presence of a charged Higgs boson. Among these processes, the radiative $b \to s(d) \gamma$ decays are of special interest, because the charged Higgs contributes to these decays at the same level as the W boson in the SM. It has already been shown that, in both the type-III and the type-C model, the inclusive $B \to X_s \gamma$ decay is very sensitive to the charged Higgs Yukawa couplings~\cite{B2Xg:2HDM:0}. Being induced by the same quark-level processes, the exclusive decay modes like $B \to K^* \gamma$ and $B \to \rho \gamma$ are also expected to be affected by these NP models. On the experimental side, especially the inclusive and exclusive decays corresponding to $b\to s\gamma$ transitions are known with good accuracy, but the branching ratios and even the direct CP and isospin asymmetries have also been measured for several $b\to d\gamma$ decays~\cite{Beringer:1900zz,Amhis:2012bh}. On the theoretical side, while the inclusive decays can be essentially calculated perturbatively with high precision, the exclusive processes are more complicated due to the interplay of non-perturbative hadronic effects~\cite{B2Vgreview}. Besides some other methods~\cite{PQCD:1,PQCD:2,PQCD:3,PQCD:4,PQCD:5,PQCD:6,PQCD:7,PQCD:8,PQCD:9,PQCD:10,PQCD:11,PQCD:12,PQCD:13,
Dimou:2012un,Lyon:2013gba}, the QCD factorization~(QCDF) approach, which will be adopted in this paper, has provided a systematic framework for the treatment of these exclusive decays~\cite{QCDF:1,QCDF:2,QCDF:3,QCDF:4,QCDF:5,QCDF:6,QCDF:7,QCDF:8}. Thanks to the experimental and theoretical improvements achieved in recent years, the exclusive $b \to s(d) \gamma$ decays are providing very important and complementary information on various NP models~\cite{Hurth:2012jn,DescotesGenon:2011yn,Mahmoudi:2009zx,Ahmady:2006yr,Ahmady:2005nc,Xiao:2003vq,
Altmannshofer:2012az,Blanke:2012tv,Lee:2006qv,Kim:2004zm,Atwood:1997zr,Jung:2012vu,Li}.

In this paper, we shall study the exclusive radiative B-meson decays in both the type-III and the type-C model. Besides the branching ratios, we shall also consider the direct CP and isospin asymmetries of these decays, which are expected to provide complementary information on the model parameters. Our paper is organized as follows: In section~2, we give a brief review on the 2HDM with MFV. In section~3, the effect of charged Higgs on $B\to V \gamma$ decays is discussed after presenting the relevant theoretical framework. In section~4, we give our detailed numerical results and discussions. We conclude in section~5.

\section{2HDM under the MFV hypothesis}
\label{sec:2HDM}

To discuss the generic 2HDM with MFV, it is convenient to rotate the two scalar doublets to the so-called ``Higgs basis", in which only one doublet~(denoted as $\Phi_1$ here) gets a nonzero VEV and behaves the same as the SM one~\cite{Davidson:2005cw}. In this basis, the Yukawa interactions of the Higgs fields with the quarks can be written as~\cite{B2Xg:2HDM:0}
\begin{align}\label{eq:Lagrangian:Yukawa}
-\mathcal{L}_Y=\bar q_L^0 \tilde \Phi_1 Y^U u_R^0
    +\bar q_L^0 \Phi_1 Y^D d_R^0
    +\bar q_L^0 \tilde\Phi_2^{(a)}T_R^{(a)}\bar Y^U u_R^0
    +\bar q_L^0       \Phi_2^{(a)}T_R^{(a)}\bar Y^D d_R^0
    +\rm{h.c.},
\end{align}
where $q_L^0$, $u_R^0$ and $d_R^0$ denote the quark fields in the interaction basis, and $\tilde\Phi_i=i\sigma_2\Phi_i^*$ with $\sigma_2$ the Pauli matrix. The $SU(3)_C$ generator $T_R^{(a)}$ acts on the quark fields and determines the colour nature of the second doublet; for a colour-singlet scalar, $T_R$ is just the identity matrix; for a colour-octet scalar, on the other hand, $T_R^{a}=T^{a}~(a=1,\cdots,8)$, denote the matrices of the fundamental representation in colour space.

The Yukawa couplings $Y^{U,D}$ and $\bar Y^{U,D}$ in eq.~(\ref{eq:Lagrangian:Yukawa}) are general complex $3 \times 3$ matrices in the quark flavour space and, under the MFV hypothesis, should have the same transformation properties in the quark flavour symmetry group $SU(3)_{Q_L}\otimes SU(3)_{U_R} \otimes SU(3)_{U_D}$. This can be achieved by requiring that the couplings $\bar Y^{U,D}$ be composed of pairs of the matrices $Y^{U,D}$~\cite{B2Xg:2HDM:0}
\begin{align} \label{eq:Y}
  \bar Y^U&=A_u^\ast (1+\epsilon_u^\ast Y^UY^{U\dagger}+\dotsc)Y^U, \nonumber\\[0.2cm]
  \bar Y^D&=A_d(1+\epsilon_d Y^UY^{U\dagger}+\dotsc)Y^D,
\end{align}
where $A_{u,d}$ and $\epsilon_{u,d}$ are generally arbitrary and complex coefficients. As discussed in ref.~\cite{B2Xg:2HDM:0}, by assuming that those involving higher powers of the Yukawa matrices are suppressed~(e.g., because they are generated at higher loops) and that the only significant deviations from proportionality between $\bar Y^{U,D}$ and $Y^{U,D}$ are due to the top-quark Yukawa couplings, one can then neglect terms involving powers of $Y^D Y^{D\dagger}$ and terms involving higher powers of $Y^U Y^{U\dagger}$, which are denoted collectively by the ellipses in eq.~(\ref{eq:Y}).

Under the assumptions for the Yukawa couplings $\bar Y^{U,D}$ made in eq.~(\ref{eq:Y}), and applying the SM unitary transformations to rotate the quark fields from the interaction to the mass-eigenstate basis, one can obtain the Lagrangian governing the Yukawa interactions between quarks and the charged Higgs boson~\cite{Davidson:2005cw,B2Xg:2HDM:0}\footnote{Since the analysis performed in ref.~\cite{B2Xg:2HDM:0} is restricted to the case of real couplings $A_u$ and $\epsilon_u$, the complex conjugate on $A_u$ and $\epsilon_u$ in eq.~\eqref{eq:Y} is unnecessary. In addition, our expression for the charged-Higgs Yukawa Lagrangian~(eq.~\eqref{eq:Lagrangian:charged Higgs}) differs from that in ref.~\cite{B2Xg:2HDM:0} by a global minus sign, which is confirmed to be a typo after communication with the authors of ref.~\cite{B2Xg:2HDM:0}. The difference has, however, no impact on the Wilson coefficients, since the vertices in eq.~\eqref{eq:Lagrangian:charged Higgs} always enter in pairs.}
\begin{align}\label{eq:Lagrangian:charged Higgs}
  \mathcal L_{H^+}=\frac{g}{\sqrt{2}m_W}\,\sum_{i,j=1}^3\,\bar u_iT_R^{(a)}(A_u^i m_{u_i}P_L-A_d^im_{d_j}P_R)V_{ij}d_jH_{(a)}^+ +\rm{h.c.},
\end{align}
where $g$ denotes the coupling constant of $SU(2)_L$ gauge group; $u_i$ and $d_j$ are the up- and down-type quark fields in the mass eigenstates, with $i,j$ the generation indices and $m_{u,d}$ the quark masses; $V$ denotes the involved CKM matrix, and $P_{R,L}=\frac{1\pm \gamma_5}{2}$ are the right- and left-handed chirality projectors. In terms of the fermion mass-eigenstate fields, the Yukawa couplings $\bar Y^{U,D}$ in eq.~(\ref{eq:Y}) can now be expressed as
\begin{align}
  A_{u,d}^i=A_{u,d}\left(1+\epsilon_{u,d}\frac{m_t^2}{v^2}\delta_{i3}\right),
\end{align}
where $v=\langle \Phi_1^0\rangle=174~{\rm GeV}$ is the VEV. Since only the couplings of charged Higgs boson to the top quark are involved for radiative $b\to s(d) \gamma$ decays, we shall drop the family index of the couplings $A_{u,d}^i$ from now on.

Following the notation used in ref.~\cite{B2Xg:2HDM:0}, we shall denote the model with a colour-singlet and the one with a colour-octet Higgs doublet as the type-III and the type-C model, respectively, both of which satisfy the principle of MFV.

It is noted that, the lepton sector, which is not discussed in this paper, is correlated with the quark sector and may affect (semi-)leptonic meson decays. It has, however, been shown explicitly in ref.~\cite{Crivellin:2012ye} that, after including contributions of the charged Higgs Yukawa interactions with leptons, the 2HDM within MFV cannot explain simultaneously the current experimental data on $R(D)$ and $R(D^{\ast})$, where $R(D^{(\ast)}) \equiv \mathcal B( B \to D^{(\ast)} \tau \nu)/\mathcal B (B  \to D^{(\ast)}\ell \nu)$~\cite{Lees:2012xj,Lees:2013uzd}.

\section{Theoretical framework for radiative B-meson decays}
\label{sec:theo}

In this section, following the analysis of refs.~\cite{QCDF:1,QCDF:2,Jung:2012vu}, we firstly present the decay rate of exclusive radiative B-meson decays, and then discuss the corrections to the Wilson coefficients due to the exchange of colourless or coloured charged Higgs boson. For more details, the readers are referred to refs.~\cite{QCDF:1,QCDF:2,QCDF:3,QCDF:4,QCDF:5,QCDF:6,QCDF:7,QCDF:8} for the former and to refs.~\cite{B2Xg:2HDM:0,B2Xg:2HDM:1,B2Xg:2HDM:2,B2Xg:2HDM:3,B2Xg:2HDM:4,B2Xg:2HDM:5,B2Xg:2HDM:6} for the latter.

\subsection{$B \to V \gamma$ decays within the QCDF framework}

Following the conventions advocated in refs.~\cite{QCDF:1,QCDF:2}, it is convenient to define the quantity
\begin{align}\label{eq:captC7}
  \mathcal C_7^{(i)}\equiv \frac{\mathcal T_\perp^{(i)}(0)}{T_1(0)}=\delta^{it}C_7^{\rm eff} + \dotsc,
\end{align}
where $i=t$ or $u$, and the subleading perturbative corrections as well as power corrections discussed in the previous subsection are denoted by the ellipses.

In terms of the quantity $\mathcal C_7^{(i)}$, the decay rate for a $\bar B \to V \gamma$ decay can be expressed as~\cite{QCDF:1,QCDF:2}
\begin{align}\label{eq:decay rate}
  \Gamma(\bar B \to V \gamma)=\frac{G_F^2}{8\pi^3}m_B^3S\left(1-\frac{m_V^2}{m_B^2}\right)^3\frac{\alpha_{\rm em}}{4\pi}m_b^2T_1(0)\left\lvert \lambda_t^{(D)}\mathcal C_7^{(t)} +\lambda_u^{(D)}\mathcal C_7^{(u)} \right\rvert^2,
\end{align}
where $S=1/2$ for $\rho^0$ and $\omega$, whereas $S=1$ for the other light vector mesons. Within the SM, the decay rate for the CP-conjugate mode $B \to \bar V \gamma$ can be obtained from eq.~(\ref{eq:decay rate}) with the replacement $\lambda_i^{(D)} \to \lambda_i^{(D)*}$. For $b\to s $ transitions, as the amplitude proportional to $\lambda_u^{(s)}$ is doubly Cabibbo suppressed, the main contribution comes from the term $\lambda_t^{(s)}\mathcal C_7^{(t)}$. However, for $b \to d $ transitions, the CKM factor $\lambda_u^{(d)}$ is of the same order as $\lambda_t^{(d)}$, and hence the corresponding amplitude cannot be neglected; indeed, interference between these two terms plays an important role in generating the CP and isospin asymmetries in these decays.

Starting with the decay rate given by eq.~(\ref{eq:decay rate}), the following three interesting observables in $B \to V \gamma$ decays can be constructed~\cite{QCDF:1,QCDF:2,QCDF:3,QCDF:4,QCDF:5,QCDF:6,QCDF:7,QCDF:8}:
\begin{itemize}
    \item The CP-averaged branching ratio
      \begin{align}
        \mathcal B (B \to V \gamma)=\tau_B \bar\Gamma(B \to V \gamma)=\tau_B\frac{\Gamma(\bar B \to V \gamma)+\Gamma(B \to \bar V \gamma)}{2},
      \end{align}
      where $\tau_B$ is the B-meson lifetime, and $\bar\Gamma$ denotes the CP-averaged decay rate. From eqs.~(\ref{eq:captC7}) and (\ref{eq:decay rate}), it can be seen that the branching ratio is proportional to $|C_7^{\rm eff}(\mu_b)|^2$ in the leading-order~(LO) approximation.

    \item The direct CP asymmetry
      \begin{align}\label{eq:directCP_asymmetry}
        \mathcal A_{CP}(B\to V\gamma)=\frac{\Gamma(\bar B \to V \gamma)-\Gamma(B \to \bar V \gamma)}{\Gamma(\bar B \to V \gamma)+\Gamma(B \to \bar V \gamma)},
      \end{align}
      which arises due to the interference between the hadronic matrix elements $\langle V \gamma\lvert \mathcal H_{\rm eff}^{(t)}\rvert \bar B\rangle$ (mainly from the operator $\mathcal O_7$) and $\langle V \gamma\lvert \mathcal H_{\rm eff}^{(u)}\rvert \bar B\rangle$~(mainly from the operator $\mathcal O_2$).

    \item The isospin asymmetries for $B \to K^* \gamma$ and $B \to \rho \gamma$ decays are defined, respectively, as
      \begin{align}\label{eq:isospin_asymmetry}
        \Delta(K^* \gamma)&=\frac{\bar\Gamma(B^0 \to K^{*0}\gamma)-\bar\Gamma(B^+ \to K^{*+}\gamma)}{\bar\Gamma(B^0\to K^{*0}\gamma)+\bar\Gamma(B^+\to K^{*+}\gamma)}, \nonumber\\[0.2cm]
        \Delta(\rho \gamma)&=\frac{\bar\Gamma(B^+\to \rho^+\gamma)}{2\bar\Gamma(B^0\to \rho^0 \gamma)}-1.
      \end{align}
      As detailed in ref.~\cite{Lyon:2013gba}, the isospin asymmetry is generated mainly from three sources: i) the weak annihilation mediated by four-quark operators, ii) the quark-loop spectator scattering through four-quark operators, and iii) the spectator scattering through chromo-magnetic operator $\mathcal O_8$. From the above definitions, it can be seen that this quantity is roughly proportional to $1/C_7^{\rm eff}(\mu_b)$ at the LO.
\end{itemize}
These observables can be used not only to test the SM but also to probe various NP beyond it~\cite{Hurth:2012jn,DescotesGenon:2011yn,Mahmoudi:2009zx,Ahmady:2006yr,Ahmady:2005nc,Xiao:2003vq,
Altmannshofer:2012az,Blanke:2012tv,Lee:2006qv,Kim:2004zm,Atwood:1997zr,Jung:2012vu,Li}. Especially, due to their different dependence on the Wilson coefficients, the information provided by these different observables is complementary to each other.

Note that in this paper we shall not discuss the indirect CP violation in the decays. The reason is that this observable remains proportional to $m_{s(d)}/m_b$ for a $b \to s(d)$ transition~\cite{PQCD:9,PQCD:12,Atwood:1997zr}, rendering it very small both within the SM and in the two 2HDMs with MFV. It is also noted that the currently available measurements are compatible with zero~\cite{Beringer:1900zz,Amhis:2012bh}.

\subsection{$B \to V \gamma$ decays in 2HDM with MFV}

For both the type-III and the type-C model introduced in section~\ref{sec:2HDM}, the tree-level FCNC transitions are highly suppressed under the MFV hypothesis, and the dominant NP contributions to $b\to s(d) \gamma$ decays arise from the photon-penguin diagrams mediated by charged Higgs boson, which contributes at the same level as the W-boson in the SM. In the approximation of vanishing strange-quark mass, these NP contributions do not generate additional operators beyond the ones present already in the SM effective Hamiltonian, and the charged-Higgs effects reside only in the short-distance Wilson coefficients at the matching scale $\mu_W$.

Using the effective Hamiltonian approach, we can obtain the matching conditions for the Wilson coefficients by matching the full theory to an effective theory with five-quark flavours, and the analytic expressions up to NLO can be found, e.g., in refs.~\cite{B2Xg:2HDM:0,B2Xg:2HDM:1,B2Xg:2HDM:2,B2Xg:2HDM:3,B2Xg:2HDM:4,B2Xg:2HDM:5,B2Xg:2HDM:6}. Especially, Degrassi and Slavich have obtained these results for the type-C model for the first time~\cite{B2Xg:2HDM:0}. It is found that, at the matching scale $\mu_W$ and up to NLO, only the Wilson coefficients $C_{4,7,8}(\mu_W)$ are affected by the charged-Higgs contributions. During the calculation of these matching coefficients, all the heavy particles~(including the top quark, vector bosons and the charged Higgs) have been integrated out simultaneously at the scale $\mu_W$, which is a reasonable approximation provided that the charged-Higgs mass $m_{H^{\pm}}$ is of the same order of magnitude as $m_{W}$ and $m_t$. The evolution of these Wilson coefficients from the matching scale $\mu_W$ down to the low-energy scale $\mu_b\simeq m_b$ remains the same as in the SM, details of which can be found, e.g., in refs.~\cite{CMM,WC:1,WC:2,WC:3,WC:4,WC:5}.

Numerically, it is found that, at the low-energy scale $\mu_b$, only the Wilson coefficients $C_{7,8}^{\rm eff}(\mu_b)$ present significant deviations from the SM predictions. In the LL approximation and taking $m_{H^\pm}=200\, \rm GeV$ and $\mu_b=2.5\, \rm GeV$, we get
\begin{align}
  C_7^{\rm eff}(\mu_b)/C_{7, \rm SM}^{\rm eff} (\mu_b)&=
  \begin{cases}
    1 -0.28 A_u^*A_d +0.045 |A_u|^2  &\text{for type-III}\\
    1 -0.27 A_u^*A_d +0.044 |A_u|^2  &\text{for type-C }
  \end{cases}
  , \label{eq:numerical C7}
\end{align}
\begin{align}
  C_8^{\rm eff}(\mu_b)/C_{8, \rm SM}^{\rm eff}(\mu_b)&=
  \begin{cases}
    1 -0.49 A_u^*A_d +0.058 |A_u|^2  & \text{for type-III}\\
    1 +0.42 A_u^*A_d -0.090 |A_u|^2  & \text{for type-C }
  \end{cases}
  , \label{eq:numerical C8}
\end{align}
where the dependence on the Yukawa couplings $A_u$ and $A_d$ has been made explicit. It can be seen that, for comparable $\lvert A_u \rvert$ and $\lvert A_d \rvert$, the dominant NP contribution comes from the term proportional to $A_u^* A_d$. Results for $C_7^{\rm eff}(\mu_b)$ are numerically quite similar for both the type-III and the type-C model and will, therefore, have indistinguishable effects on the branching ratios. There is, however, one main difference between the type-III and the type-C model, \textit{i.e.}, their contributions to $C_8^{\rm eff}(\mu_b)$ have similar magnitudes but with opposite signs, which will affect predictions for the direct CP and isospin asymmetries in exclusive radiative B-meson decays. It is therefore expected that these interesting observables may have the potential to distinguish between the two models, as will be detailed in the next section.

It is noted that only the charged-Higgs contributions are relevant for the radiative B-meson decays. With the parametrization of eq.~(\ref{eq:Lagrangian:charged Higgs}), the charged-Higgs Yukawa interactions with quarks in the type-III model are the same as the ones in the A2HDM~\cite{Jung:2012vu}, with the replacements $ A_{u(d)} \leftrightarrow \varsigma_{u(d)}$. Therefore, constraints on the model  parameters and correlations between the various observables are the same in these two models.

\section{Numerical results and discussions}

With the theoretical framework recapitulated in the previous section, we proceed to present our numerical results and discussions in this section.

\subsection{SM predictions and experimental data}

%%%%%%%%%%%%%%%%%%%%%%%%%%%%%%%%%%%%%%%%%%%%%%%%%%%%%%%%%%%%%%%%%%%
\begin{table}[t]
  \centering
  \begin{tabular}{|c|}
    \hline
  \tabcolsep 0.08in
  \begin{tabular}{c l l|c l c}
    $\alpha_{\rm e}$   & $1/137.036$                    & \cite{Beringer:1900zz}       &  $\sin^2\theta_W$   &  $0.23146 \pm 0.00012$                               & \cite{Beringer:1900zz}   \\
    $\alpha_s(m_Z)$   & $0.1184\pm 0.0007$       &  \cite{Beringer:1900zz}      &   $m_Z$                    &  $(91.1876\pm 0.0021)\, \rm{GeV}$              & \cite{Beringer:1900zz}   \\
    $|V_{us}|$              & $0.2252 \pm 0.0009$      &  \cite{Beringer:1900zz}     &  $m_W$                    &  $(80.385\pm 0.015) \, \rm{GeV}$                 & \cite{Beringer:1900zz}   \\
    $|V_{ub}|$              & $0.00415 \pm 0.00049$  &  \cite{Beringer:1900zz}     &   $m_b(m_b)$            &  $(4.18\pm 0.03)\, \rm{GeV}$                        & \cite{Beringer:1900zz}   \\
    $|V_{cb}|$              & $0.0409 \pm 0.0011 $     &  \cite{Beringer:1900zz}     &   $m_c(m_c)$             &  $(1.275\pm 0.025) \, \rm{GeV}$                   & \cite{Beringer:1900zz}   \\
    $\gamma$           & $(68_{-11}^{+10})^\circ$       &  \cite{Beringer:1900zz}     &   $m_t^{\rm pole}$      &  $(173.18\pm 0.94)\, \rm {GeV}$   & \cite{mt}     \\
    $\lambda_{B,+}(1\,{\rm GeV})$ & $(0.460\pm 0.110)\, \rm{GeV}$ & \cite{Braun:2003wx} & $f_{B_s}$
    &  $(227.6\pm 5.0)\, \rm {MeV}$   & \cite{FF:Bs:1,FF:Bs:2,FF:Bs:3,Laiho:2009eu} \\
  \end{tabular}\\
    \hline
  \tabcolsep 0.14in
  \begin{tabular}{c c c c c c}
    Meson     &$a_{1,\parallel}(1\,{\rm GeV})$& $a_{1,\perp}(1\,{\rm GeV})$&$a_{2,\parallel}(1\,{\rm GeV})$&$a_{2,\perp}(1\,{\rm GeV})$&ref.\\
    $\rho$    & ---					 & ---			             & $0.15\pm 0.07$                 & $0.14\pm 0.06$ &\cite{Gegenbauer:rho:1,Gegenbauer:rho:2}\\
    $\bar K^\ast$  & $0.03\pm 0.02$	         & $0.04\pm 0.03$	             & $0.11\pm 0.09$	           & $0.10\pm 0.08$ & \cite{Gegenbauer:KV:1,Gegenbauer:KV:2,Gegenbauer:KV:3}\\
   $\phi$	   & ---					 & ---				     & $0.18\pm 0.08$		    & $0.14\pm 0.07$& \cite{Gegenbauer:phi}\\
   $\omega$& ---					 & ---				     & $0.15\pm 0.14$		    & $0.14\pm 0.12$& \cite{Jung:2012vu}\\
  \end{tabular}\\
    \hline
  \tabcolsep 0.18in
  \begin{tabular}{cc c c c}
  Meson 		& $f_\parallel/{\rm GeV}$ 	& $f_\perp(2~{\rm GeV})/f_\parallel$	& $T_1(0)$	            & ref.\\
  $\rho$		& $0.215\pm 0.006$	& $0.70\pm 0.04$				& $0.276\pm 0.039$   & \cite{Beringer:1900zz,PQCD:9}\\
  $K^\ast$     	& $0.209\pm 0.007$       & $0.73\pm 0.04$				& $0.302\pm 0.052(B_{u,d})$   & \cite{Beringer:1900zz,PQCD:9}\\
			&					&							& $0.274\pm 0.045(B_s)\:\:\:$   & \\
  $\phi$		& $0.229\pm 0.003$	& $0.75\pm 0.02$				& $0.335\pm 0.043$   & \cite{Beringer:1900zz,FF:phi}\\
  $\omega$	& $0.188\pm 0.010$	& $0.70\pm 0.10$				& $0.237\pm 0.055$   & \cite{Beringer:1900zz,Jung:2012vu}\\
  \end{tabular}\\
    \hline
  \end{tabular}
  \caption{\small The relevant input parameters used in our numerical analysis. The meson masses and lifetimes can be found in ref.~\cite{Beringer:1900zz}. The tensor form factors $T_1(0)$ are given for $B_{u,d}\to K^*,\rho,\omega$ and $B_s\to \bar{K}^*,\phi$. The ratio $f_{B_s}/f_{B_d}=1.201\pm 0.017$~\cite{Laiho:2009eu} is used to determine the decay constant $f_{B_d}$. Details about how to obtain these hadronic input parameters could be found in ref.~\cite{Jung:2012vu}.}
  \label{tab:input}
\end{table}
%%%%%%%%%%%%%%%%%%%%%%%%%%%%%%%%%%%%%%%%%%%%%%%%%%%%%%%%%%%%%%%%%%%

With the input parameters collected in table~\ref{tab:input}, our SM predictions for the observables in exclusive radiative B-meson decays are listed in table~\ref{tab:SM}, in which the theoretical uncertainties are obtained by varying each input parameter within its respective range and adding the individual uncertainty in quadrature. For sake of completeness, we also present in table~\ref{tab:SM} our results for the branching ratios and direct CP asymmetries of inclusive $B\to X_{s,d} \gamma$ decays, the theoretical framework of which could be found, e.g., in refs.~\cite{B2Xg:BR:1,B2Xg:BR:2,B2Xg:BR:3,B2Xg:BR:4} and \cite{B2Xg:CP:1,B2Xg:CP:2,B2Xg:CP:3,B2Xg:CP:4}, respectively. All the experimental data is taken from the Heavy Flavor Averaging Group~\cite{Amhis:2012bh}.

%%%%%%%%%%%%%%%%%%%%%%%%%%%%%%%%%%%%%%%%%%%%%%%%%%%%%%%%%%%%%%%%%%%
\begin{table}[t]
  \centering
  \tabcolsep 0.1in
  \begin{tabular}{|l l l|l l l|}
    \hline
    Observable & SM & Exp. & Observable & SM & Exp.\\
    \hline
    $\mathcal B(B \to X_s \gamma)$& $316_{-27}^{+26}$ &
    $343\pm 22$&
    $\mathcal A_{CP}(B \to X_s \gamma)$&$2.60_{-3.33}^{+0.78}$&$-0.8\pm 2.9$\\
    $\mathcal B(B^+ \to K^{*+}\gamma)$&$42.4_{-15.0}^{+17.3}$&$42.1\pm
    1.8$&
    $\mathcal A_{CP}(B^+ \to
    K^{*+}\gamma)$&$0.38_{-0.26}^{+0.32}$&$18\pm 29$\\
    $\mathcal B(B^0\to K^{*0}\gamma)$&$42.6_{-14.7}^{+16.8}$&$43.3 \pm
    1.5$&
   $\mathcal A_{CP}(B^0\to
    K^{*0}\gamma)$&$0.74_{-0.25}^{+0.30}$&$-0.7\pm 1.9$\\
    $\mathcal B (B_s \to \phi \gamma)$&$53.7_{-15.2}^{+16.7}$&$35\pm 3$&
    $\mathcal A_{CP}(B_s \to \phi \gamma)$& $0.52_{-0.15}^{+0.20}$&---\\
\hline
    $\mathcal B(B \to X_d \gamma)$&
    $15.2_{-3.8}^{+3.7}$& $14.1\pm 4.9$&
    $\mathcal A_{CP}(B \to X_d \gamma)$&$-57.4_{-17.3}^{+73.6}$&---\\
    $\mathcal B(B^+ \to \rho^+
    \gamma)$&$1.64_{-0.55}^{+0.60}$&$0.98_{-0.24}^{+0.25}$&
    $\mathcal A_{CP}(B^+ \to \rho^+
    \gamma)$&$-12.2_{-5.2}^{+3.4}$&$-11\pm 33$\\
    $\mathcal B(B^0 \to \rho^0
    \gamma)$&$0.84_{-0.30}^{+0.32}$&$0.86_{-0.14}^{+0.15}$ &
    $\mathcal A_{CP}(B^0 \to \rho^0
    \gamma)$&$-12.0_{-4.4}^{+3.3}$&---\\
    $\mathcal B(B^0 \to \omega
    \gamma)$&$0.62_{-0.29}^{+0.35}$&$0.44_{-0.16}^{+0.18}$&
    $\mathcal A_{CP}(B^0 \to \omega\gamma)$&$-11.6_{-4.5}^{+3.4}$&---\\
    $\mathcal B (B_s \to \bar K^{0*}\gamma)$&$1.71_{-0.65}^{+0.72}$&---&
    $\mathcal A_{CP} (B_s \to \bar
    K^{0*}\gamma)$&$-11.3_{-4.3}^{+3.2}$&---\\
\hline
    $\Delta (K^* \gamma)$&$4.2_{-2.5}^{+2.4}$&$5.2\pm 2.6$&
    $\Delta (\rho \gamma)$&$-9.5_{-6.9}^{+8.6}$&$-46_{-16}^{+17}$\\
    \hline
  \end{tabular}
  \caption{\small SM predictions and experimental measurements for the observables in radiative B-meson decays. The branching ratios are given in units of $10^{-6}$, and the direct CP and isospin asymmetries in units of $10^{-2}$. For the inclusive $B \to X_{s,d} \gamma$ decays, the values given here correspond to a photon energy cut at $E_\gamma=1.6\,\rm{GeV}$~\cite{Amhis:2012bh}.}
  \label{tab:SM}
\end{table}
%%%%%%%%%%%%%%%%%%%%%%%%%%%%%%%%%%%%%%%%%%%%%%%%%%%%%%%%%%%%%%%%%%%

It is observed that our predictions for the branching ratios of inclusive $B\to X_{s}\gamma$ and $B\to X_{d}\gamma$ decays are in good agreement with the experimental measurements, implying very stringent constraints on various NP models~\cite{Hurth:2012jn,DescotesGenon:2011yn,Mahmoudi:2009zx,Ahmady:2006yr,Ahmady:2005nc,Xiao:2003vq,
Altmannshofer:2012az,Blanke:2012tv,Lee:2006qv,Kim:2004zm,Atwood:1997zr,Jung:2012vu,Li}. For the direct CP asymmetries, on the other hand, due to the appearance of long-distance effect in the interference
of the electro-magnetic dipole amplitude with the amplitude for an up-quark penguin transition
accompanied by soft gluon emission~(the so-called \textit{``resolved photon contributions"}), there are still quite large uncertainties in the theoretical predictions, lowering the predictive power of these observables~\cite{B2Xg:CP:4}.

For the exclusive $B\to V\gamma$ decays, within the QCDF formalism, the main theoretical uncertainties stem from the hadronic input parameters and the variation of renormalization scale $\mu_b$. We have also added an additional $15\%$ global uncertainty in all exclusive observables to account for the non-factorizable effects, which have not yet been included in the QCDF framework. It is noted that, taking into account their respective uncertainties, our predictions for these observables are generally in good agreement with the current data, except for one tension observed for $\Delta (\rho \gamma)$, which has however rather large experimental errors. Thus, stringent constraints on the two 2HDMs with MFV are expected from these exclusive decays.

\subsection{Procedure in numerical analysis}

As shown explicitly in eqs.~(\ref{eq:numerical C7}) and (\ref{eq:numerical C8}), the relevant model parameters in our case can be chosen as the Yukawa couplings $|A_u|$ and $|A_u^*A_d|$, the phase $\theta$~(defined as $A_u^*A_d=|A_u^*A_d|e^{-i\theta}$), as well as the charged-Higgs mass $m_{H^\pm}$. As detailed in ref.~\cite{B2Xg:2HDM:0}, a stringent upper bound on the coupling $\lvert A_u \rvert$ can be obtained from the process $Z\to b \bar b$. Limits on the charged-Higgs mass from flavour observables and direct searches, however, depend strongly on the assumed Yukawa structure. The latest bound on the type-II 2HDM from $\mathcal B(B\to X_s \gamma)$ gives $m_{H^\pm} \ge 380~{\rm GeV}$ at $95\%$ confidence level~(C.L.)~\cite{B2Xg:2HDM:6}. Within the A2HDM, on the other hand, it is still possible to have a light charged Higgs~\cite{Jung:2012vu,Jung:2010ik,Jung:2010ab,Celis:2012dk,Celis:2013rcs,Jung:2013hka}. Assuming that the charged Higgs decays only into fermions $u_i\bar d_j$ and $l^+\nu_l$, LEP established the limit $m_{H^\pm}> 78.6~{\rm GeV}$~($95\%$ C.L.)~\cite{Searches:2001ac}, which is independent of the Yukawa structure. A charged Higgs produced via top-quark decays has also been searched for at Tevatron~\cite{Abulencia:2005jd,Abazov:2009aa} and LHC~\cite{Aad:2012tj,Chatrchyan:2012vca}; these searches are, however, not readily translatable into constraints for the model parameters considered here. With these points kept in mind, we shall restrict the model parameters in the following ranges:
\begin{equation}
|A_u|\in [0,3], \quad |A_u^*A_d|\in[0,60], \quad \theta\in[-180^\circ,180^\circ], \quad m_{H^\pm}\in[80,500]\,\rm GeV.
\end{equation}

The colored scalars in the 2HDM within MFV may also alter the production and/or decay rates of the neutral Higgs boson $h$ discovered at the LHC, since the couplings of $h$ with gluons, photons and $Z\gamma$ may be affected by the colored scalar-mediated loops. However, these 2HDM contributions arise from the triple- and quartic-scalar interactions present in the Higgs potential and are, therefore, independent of the Yukawa interactions discussed in this paper~\cite{Wise}. For the phenomenological implications of colored scalars at the LHC, the readers are referred to refs.~\cite{Wise,Gresham:2007ri,Gerbush:2007fe,Burgess:2009wm,Idilbi:2010rs,Cao:2013wqa}.

In order to derive the allowed parameter space from radiative B-meson decays, we adopt the same procedure as in ref.~\cite{Jung:2012vu}: each point in the parameter space corresponds to a theoretical range, constructed as the prediction for an observable in that point together with the corresponding theoretical uncertainty. If this range has overlap with the $2\sigma$ range of the experimental data, the point is regarded as allowed. To incorporate the theoretical uncertainty, we use the statistical treatment based on frequentist statistics and Rfit scheme~\cite{Hocker:2001xe}, which has been implemented in the CKMfitter package~\cite{Charles:2004jd}. Here the basic observation is that, while the experimental data approximatively yield a Gaussian distribution of an observable, a theoretical calculation does not. The latter depends on a set of input parameters like form factors, decay constants and Gegenbauer moments etc., for which no probability distribution is known. The Rfit scheme assumes no particular distribution for the theory parameters, only that they are constrained to certain allowed ranges with an equal weighting, irrespective of how close they are from the edges of the allowed range. In addition, for simplicity, the relative theoretical uncertainty is assumed to be constant at each point in the parameter space. This is a reasonable assumption, since the main theoretical uncertainties are due to the hadronic input parameters, common to both the SM and the NP contributions. Therefore, the theoretical range for an observable at each point in the parameter space is obtained by varying each input parameter within its respective allowed range and then adding the individual uncertainty in quadrature.

As can be seen from table~\ref{tab:SM}, at present most of the observables in radiative B-meson decays have not been precisely measured and/or their theoretical predictions are still quite uncertain. It is therefore interesting to investigate the correlations between these various observables, which might be helpful to gain further insights into the model parameters, with improved experimental measurements and theoretical predictions expected in the near future. Since both the experiment~\cite{Amhis:2012bh} and theory~\cite{B2Xg:BR:1,B2Xg:BR:2} have acquired a precision of a few percent for the branching ratio $\mathcal B (B \to X_s \gamma)$, we shall explore these correlations within the allowed parameter space constrained by this observable. For simplicity, we do not consider the theoretical uncertainty at each point in the parameter space. As the theoretical uncertainties of the other observables are mostly independent from the one of $B\to X_s\gamma$ and are approximately common to both the SM and the NP contributions, the cross for the SM uncertainties shown in the plots is also applied to each of these points.

\subsection{$B\to X_{s,d}\gamma$ decays in 2HDM with MFV}

%%%%%%%%%%%%%%%%%%%%%%%%%%%%%%%%%%%%%%%%%%%%%%%%%%%%%%%%%%%%%%%%%%%
\begin{figure}[htbp]
  \centering
  \subfigure[real couplings]{\includegraphics[width=0.91\textwidth]{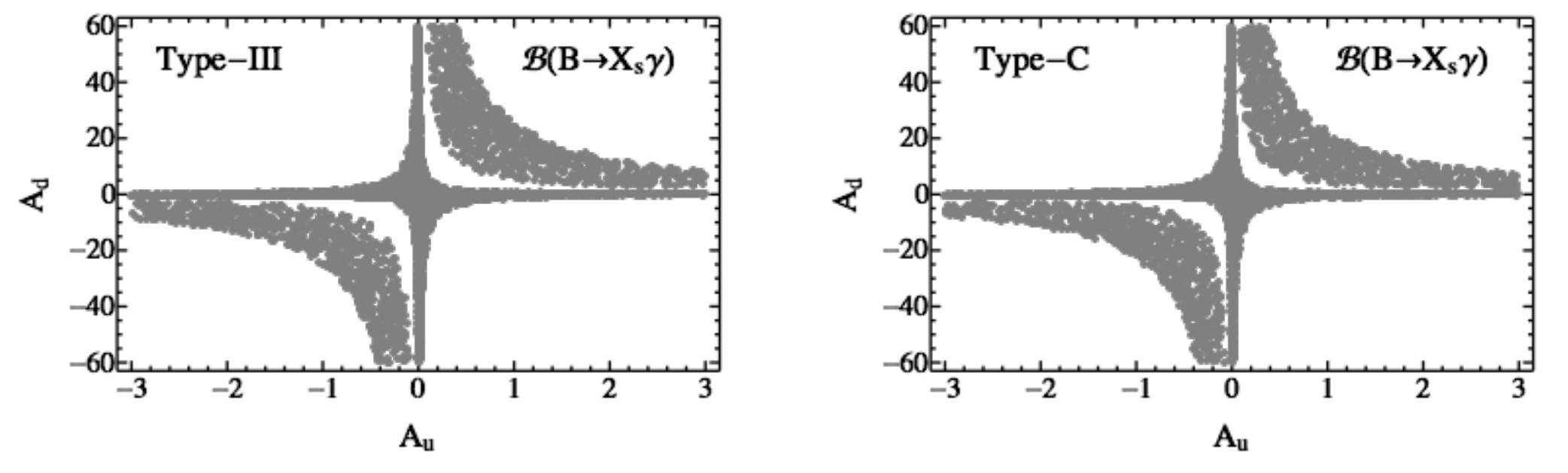}}
  \subfigure[complex couplings]{\includegraphics[width=0.92\textwidth]{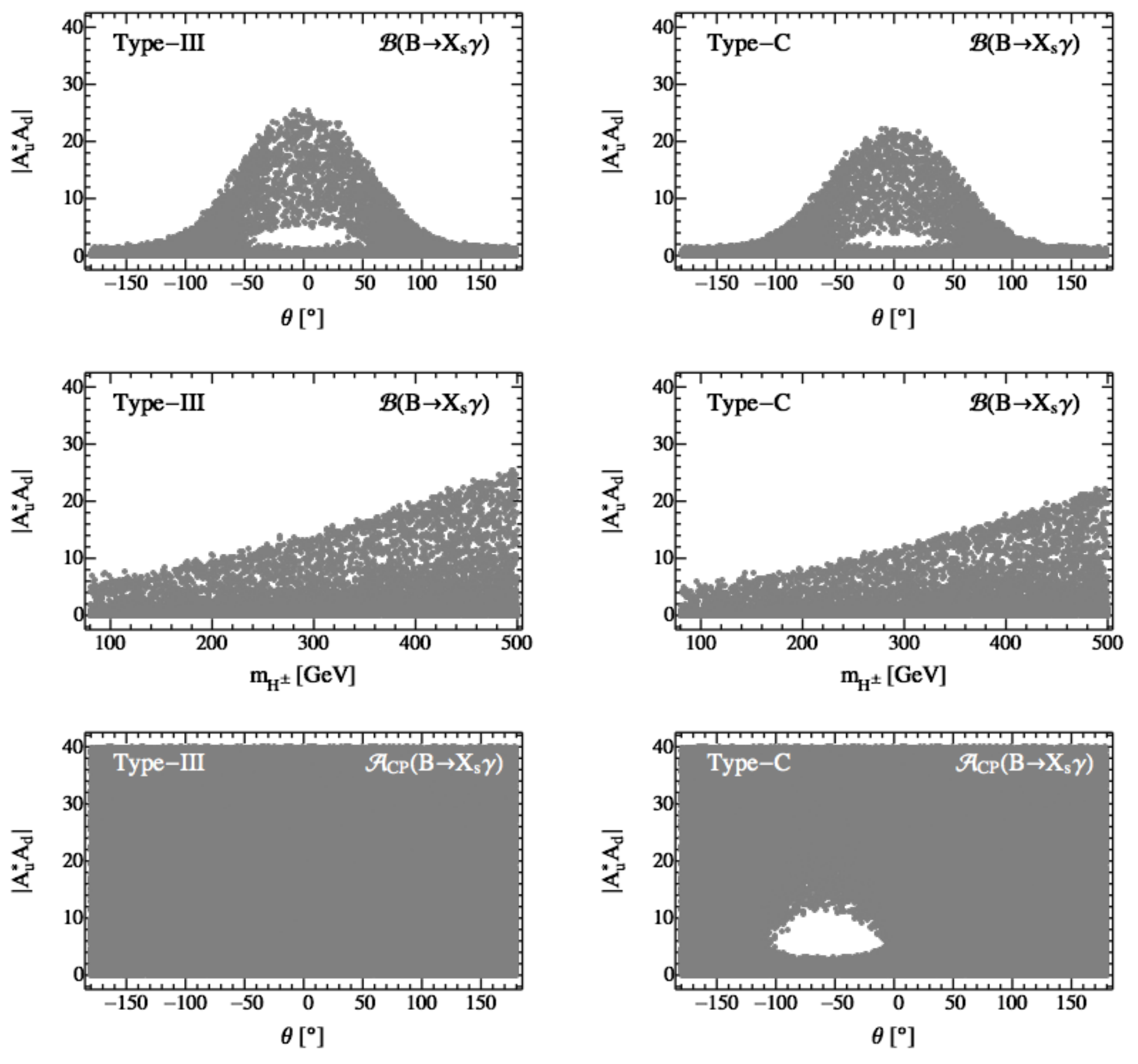}}
  \caption{\small Constraints on the couplings $A_u$ and $A_d$ for type-III and type-C models from $B \to X_s\gamma$, plotted in the planes $A_u-A_d$~(real couplings), $|A_u^*A_d|-\theta$ and $|A_u^*A_d|-m_{H^\pm}$~(complex couplings).}
  \label{fig:B2Xsg}
\end{figure}
%%%%%%%%%%%%%%%%%%%%%%%%%%%%%%%%%%%%%%%%%%%%%%%%%%%%%%%%%%%%%%%%%%%

The branching ratios of $B \to X_{s,d} \gamma$ decays are, at the LO approximation, also proportional to $\lvert C_7^{\rm eff}(\mu_b) \rvert^2$, for which the dominant NP contribution comes from the term proportional to $A_u^*A_d$. Consequently, a stringent constraint on the combination $A_u^*A_d$ is expected from these observables. This is exemplified in figure~\ref{fig:B2Xsg}, in which we show constraints on the couplings $A_u$ and $A_d$ for both the type-III and the type-C model from the $B \to X_s\gamma$ decay, plotted in the planes $A_u-A_d$~(real couplings), $|A_u^*A_d|-\theta$ and $|A_u^*A_d|-m_{H^\pm}$~(complex couplings). Constraints from the $B \to X_d \gamma$ decay are similar but slightly weaker, and hence not shown here. From these plots, the following observations are made:
\begin{itemize}
      \item For both the real and complex cases, constraints from the branching ratio are almost indistinguishable for the type-III and the type-C model. This can be understood because the branching ratio is proportional to $|C_7^{\rm eff}(\mu_b)|^2$ to the first order, for which there are no significant differences between these two models~(see eq.~(\ref{eq:numerical C7})).

      \item In the case of real couplings, there exist two allowed regions under the constraint of $\mathcal B(B \to X_s \gamma)$. The region close to the axes corresponds to the case where the NP contribution is small and constructive with the SM one. In the other region, on the other hand, simultaneously large and same-sign values for $A_u$ and $A_d$ are allowed, corresponding to the case where the interference becomes destructive and makes the coefficient $C_7^{\rm eff}(\mu_b)$ sign-flipped. However, the regions with simultaneously large values for $A_u$ and $A_d$ but with opposite signs are already excluded.

      \item In the case of complex couplings, the interference between the SM and NP contributions depends on the phase $\theta$. Especially, for $\theta\approx \pm 180^\circ$ only a small region with smaller $|A_u^*A_d|$ remains due to the constructive effect between them; while for $\theta \approx 0^\circ$, the interference becomes destructive and there exist two allowed regions, corresponding respectively to the case with relatively small NP influence~(the lower region) and the case where the NP contribution is about twice the size of the SM one~(the upper region). In addition, the combination $|A_u^* A_d|$ is strongly correlated with the charged-Higgs mass, with large values only allowed for large $m_{H^{\pm}}$.

      \item Since the experimental data and the theoretical prediction for the direct CP asymmetry $\mathcal A_{CP}(B \to X_s \gamma)$ still suffer large uncertainties, this observable gives almost no constraint on the model parameters, except for the small excluded region in the type-C model.
\end{itemize}

%%%%%%%%%%%%%%%%%%%%%%%%%%%%%%%%%%%%%%%%%%%%%%%%%%%%%%%%%%%%%%%%%%%
\begin{figure}[htbp]
  \centering
  \includegraphics[width=0.92\textwidth]{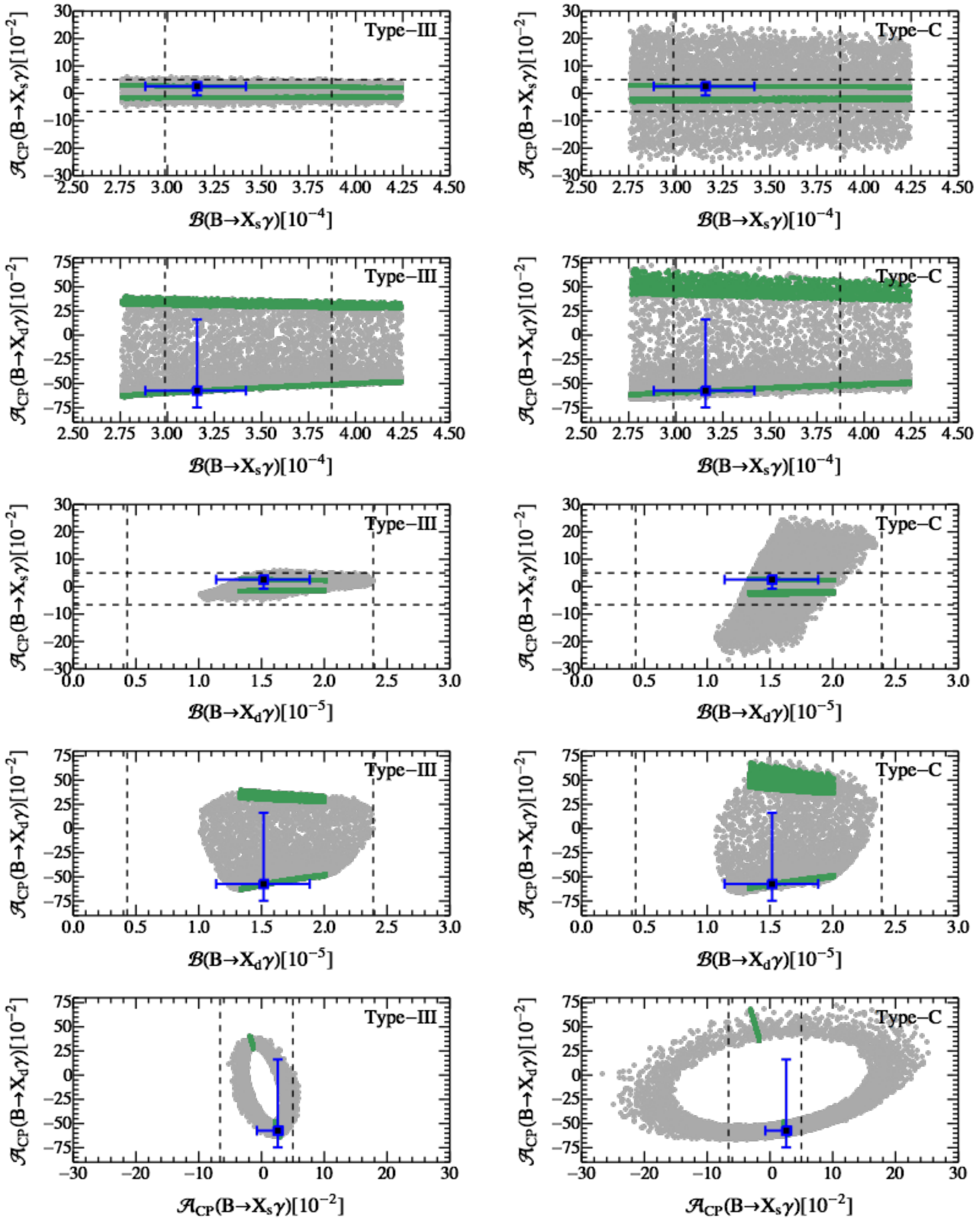}
  \caption{\small Correlation plots between the observables of $B\to X_{s,d} \gamma$ decays. The allowed regions are shown by green~(dark grey) and grey~(light grey) points, which are obtained in the case of real and complex couplings, respectively. The dashed lines denote the experimental data with $2\sigma$ range, while the SM predictions with the corresponding theoretical range are shown by the blue~(dark) cross. This cross is also applied to each of the points to account for the theoretical uncertainty at that point.}
  \label{fig:corr:B2Xg}
\end{figure}
%%%%%%%%%%%%%%%%%%%%%%%%%%%%%%%%%%%%%%%%%%%%%%%%%%%%%%%%%%%%%%%%%%%

Since the branching ratio $\mathcal B(B \to X_s \gamma)$ is a key observable, it is interesting to investigate its correlations with the other observables. Furthermore, under the constraints of $\mathcal B(B \to X_s \gamma)$, correlations between the other observables are also expected to be significantly affected. These are shown in figure~\ref{fig:corr:B2Xg} for the inclusive $B\to X_{s,d} \gamma$ decays in both the type-III and the type-C model, from which one observes only mild correlations for most observables, the exception being the correlation between the two direct CP asymmetries.

It is also interesting to note that the NP contributions exhibit large and small deviations from the SM predictions for $\mathcal A_{CP}(B \to X_s \gamma)$ and $\mathcal A_{CP}(B \to X_d \gamma)$, respectively. The predicted ranges are similar for $\mathcal A_{CP}(B \to X_d \gamma)$ but quite different for $\mathcal A_{CP}(B \to X_s \gamma)$ in the two models; especially, the type-C model gives much wider ranges for $\mathcal A_{CP}(B \to X_s \gamma)$ than the ones predicted in the type-III model. To understand these observations, we should firstly recall that the dominant NP effects in our case are encoded in the two dipole coefficients $C_7^{\rm eff}$ and $C_8^{\rm eff}$, and the main contributions to these observables consist of three terms proportional respectively to $\text{Im}\left(C_2/C_7^{\rm eff}\right)$, $\text{Im}\left(C_8^{\rm eff}/C_7^{\rm eff}\right)$ and $\text{Im}\left[\lambda_u^{(D)}/\lambda_t^{(D)} \cdot C_2/C_7^{\rm eff}\right]$, see eq.~(12) in ref.~\cite{B2Xg:CP:4} for details. As the SM contributions to $C_2$ and $C_{7,8}^{\rm eff}$ are all real, only the last term contributes and makes a difference between $\mathcal A_{CP}(B \to X_s \gamma)$ and $\mathcal A_{CP}(B \to X_d \gamma)$, being doubly Cabibbo suppressed for the former but absent of this suppression for the latter. In the 2HDMs considered here, on the other hand, the charged-Higgs Yukawa couplings are generally complex and, therefore, all the three terms can contribute and do not suffer any CKM-suppression. Thus, compared to the SM result, a wider region of $\mathcal A_{CP}(B \to X_s \gamma)$ is predicted in the type-C model due to the constructive interference from the NP contributions; in the type-III model, the NP contributions to $C_8^{\rm eff}$ have an opposite sign, which makes the overall NP effect being destructive and leading to a direct CP asymmetry with small derivations from its SM prediction. For $\mathcal A_{CP}(B \to X_d \gamma)$, however, the NP contributions are not doubly Cabibbo enhanced relative to the SM ones, which results in only small deviations between them. Consequently, the observable $\mathcal A_{CP}(B \to X_s \gamma)$ is more suitable to distinguish between the two models.

As a final comment, we should note that the large deviations from the SM value for $\mathcal A_{CP}(B \to X_s \gamma)$ are, however, not favored by the current experimental data, part of which corresponds to the excluded region shown in the last plot of figure~\ref{fig:B2Xsg}. Further insights into the model parameters provided by these observables have to be complemented by the improved experimental precision and theoretical progress expected in the near future.

\subsection{$B \to K^* \gamma$ decays in 2HDM with MFV}
\label{sec:B2KV}

For the exclusive $B \to V \gamma$ decays, although the decay amplitudes are proportional to the coefficient $C_{7}^{\mathrm{eff}}(\mu_b)$ through the quantity ${\cal C}_7^{(i)}$ defined in eq.~(\ref{eq:captC7}), predictions for the branching ratios still suffer large uncertainties mainly due to the tensor form factor $T_1(0)$. Accordingly, the branching ratios of exclusive decays could not provide further constraints on the model parameters with respect to that of the inclusive $B \to X_s \gamma$ decay. However, another two interesting observables, the direct CP and isospin asymmetries, can be constructed for the exclusive modes, both of which show a different dependence on the NP parameters from that of the branching ratios. Thus, different constraints on the model parameters are expected from these two observables. In this subsection, we shall firstly discuss $B\to K^*\gamma$ decays.

%%%%%%%%%%%%%%%%%%%%%%%%%%%%%%%%%%%%%%%%%%%%%%%%%%%%%%%%%%%%%%%%%%%
\begin{figure}[htbp]
  \centering
  \subfigure[real couplings]{\includegraphics[width=0.91\textwidth]{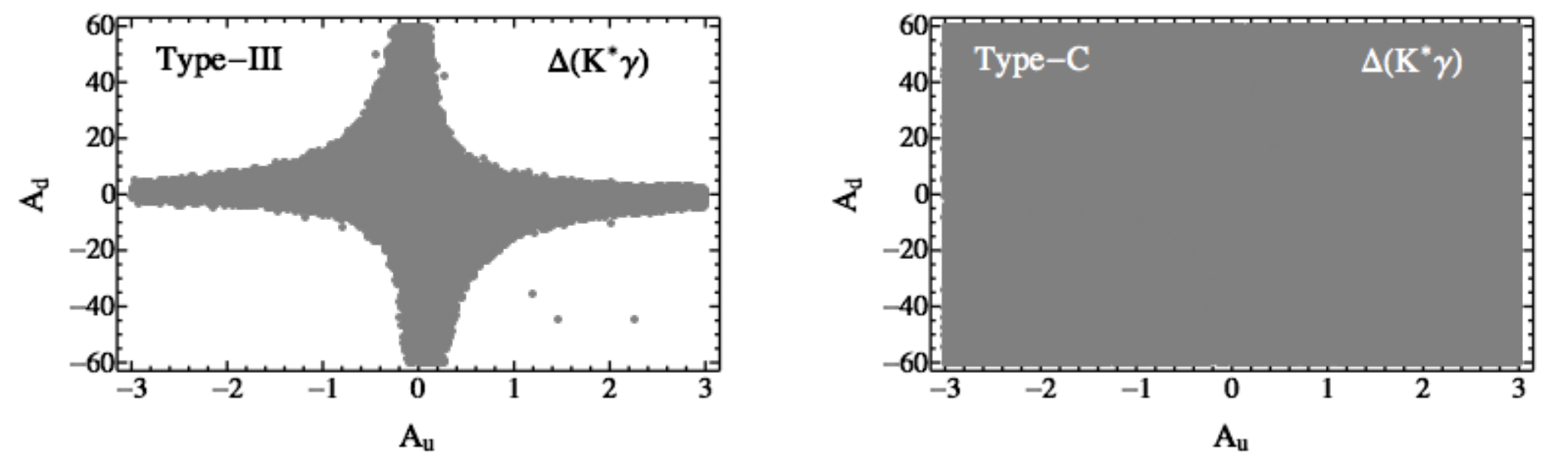}}
  \subfigure[complex couplings]{\includegraphics[width=0.92\textwidth]{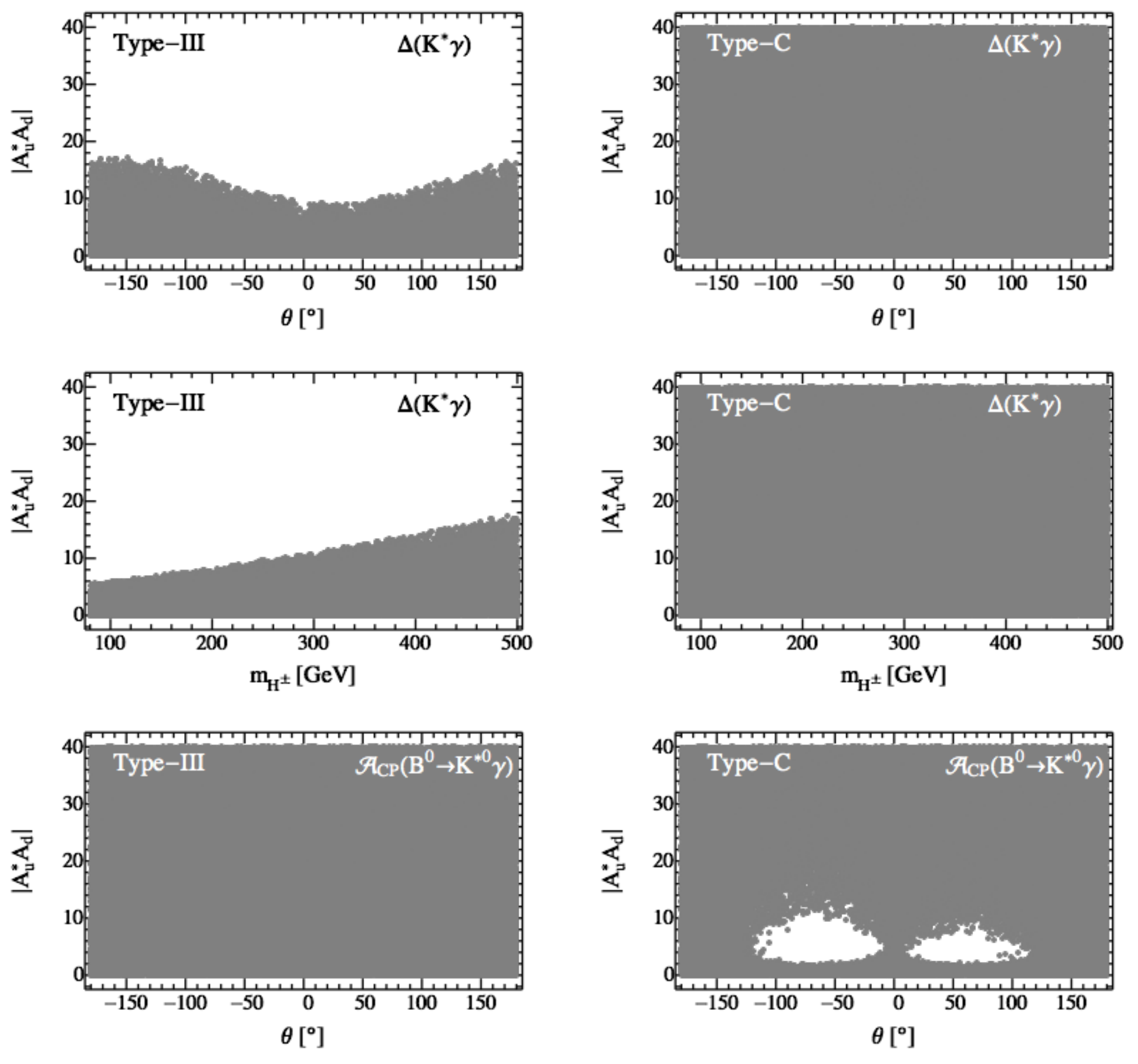}}
  \caption{\small Constraints on the couplings $A_u$ and $A_d$ from the observables $\Delta(K^*\gamma)$ and $\mathcal A_{CP}(B^0 \to K^{*0} \gamma)$. The other captions are the same as in figure~\ref{fig:B2Xsg}.}
  \label{fig:KV}
\end{figure}
%%%%%%%%%%%%%%%%%%%%%%%%%%%%%%%%%%%%%%%%%%%%%%%%%%%%%%%%%%%%%%%%%%%

Under the constraints of the current experimental data on $B\to K^*\gamma$ decays, we show in figure~\ref{fig:KV} the allowed regions of the charged-Higgs Yukawa couplings $A_{u}$ and $A_d$ for both the type-III and type-C model, plotted also in the planes $A_u-A_d$~(real couplings), $|A_u^*A_d|-\theta$ and $|A_u^*A_d|-m_{H^\pm}$~(complex couplings). From these plots, we make the following observations:
\begin{itemize}
      \item In the case of type-III model, constraints derived from the isospin asymmetry $\Delta(K^*\gamma)$ exhibit a different dependence on the NP phase $\theta$ with respect to that of the branching ratio shown in figure~\ref{fig:B2Xsg}. This is caused by the different dependence on the coefficient $C_{7}^{\mathrm{eff}}(\mu_b)$ between these observables; while the branching ratios are proportional to $|C_{7}^{\mathrm{eff}}(\mu_b)|^2$, the isospin asymmetry varies like $1/C_{7}^{\mathrm{eff}}(\mu_b)$, both being at the LO approximation. Furthermore, the large same-sign solutions allowed by $\mathcal B(B \to X_s \gamma)$ are already excluded once constraints from $\Delta(K^*\gamma)$ are taken into account.

      \item In the case of type-C model, on the other hand, the isospin asymmetry puts almost no bounds on the Yukawa couplings; especially, it could not exclude the regions with large $|A_u|$ and $|A_d|$. This is quite different from that observed in the type-III model. To understand this, we should note that, among the three sources of the isospin asymmetry pointed out below eq.~(\ref{eq:isospin_asymmetry}), the contribution from the spectator scattering through the chromo-magnetic operator $\mathcal O_8$~(i.e., the terms proportional to ${\rm Re}(C_7^{\rm eff *}C_8^{\rm eff})$ and $|C_8^{\rm eff}|^2$) is more relevant in the regions with large $|A_u|$ and $|A_d|$. In this region, the isospin asymmetry will be dominated by these two terms and, since the predicted $C_8^{\rm eff}(\mu_b)$ have similar magnitudes but opposite signs, the different interference between them results in numerical values with opposite signs between these two models. Taking into account the current experimental constraint with $2\sigma$ error bars, $0\leq\Delta(K^*\gamma)\leq10.4$, we can therefore exclude the case in which the predicted $\Delta(K^*\gamma)$ lies outside this range, which corresponds to the type-III model discussed above. The type-C model belongs to the other case, in which the predicted $\Delta(K^*\gamma)$ is still consistent with the experimental measurement, and hence the regions with large $|A_u|$ and $|A_d|$ still could not be excluded.

      \item Since there are still large theoretical and experimental uncertainties for the direct CP asymmetries in these decays, again almost no constraints can be obtained from these observables.  For the type-C model, however, some regions with small $|A_u^*A_d|$ have already been excluded by the observable $\mathcal A_{CP}(B^0 \to K^{*0} \gamma)$. This is due to the same reason as explained in the case of inclusive $B\to X_s \gamma$ decay.

      \item The branching ratio and the isospin asymmetry are very complementary to each other. Although the constraints from exclusive branching ratios are slightly weaker than the ones from $\mathcal B(B \to X_s \gamma)$, their combinations with the isospin asymmetry play an important role in further reducing the allowed parameter space. This will be explored in section~\ref{sec:concl}.

\end{itemize}

%%%%%%%%%%%%%%%%%%%%%%%%%%%%%%%%%%%%%%%%%%%%%%%%%%%%%%%%%%%%%%%%%%%
\begin{figure}[htbp]
  \centering
  \includegraphics[width=0.98\textwidth]{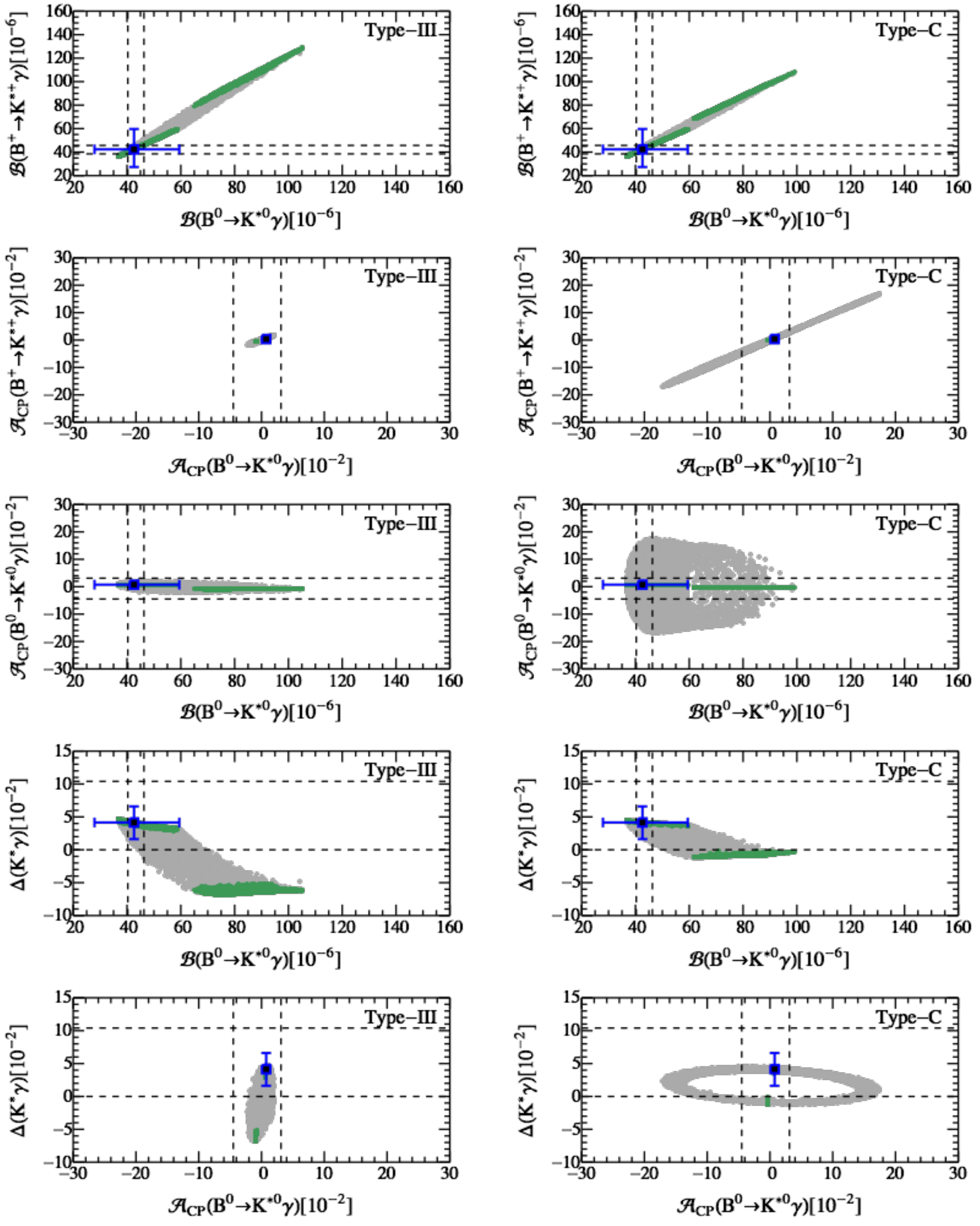}
  \caption{\small Correlation plots between the observables of $B \to K^* \gamma$ decays within the parameter space allowed by $\mathcal B(B \to X_s \gamma)$. The other captions are the same as in figure~\ref{fig:corr:B2Xg}.}
  \label{fig:corr:B2KVg}
\end{figure}
%%%%%%%%%%%%%%%%%%%%%%%%%%%%%%%%%%%%%%%%%%%%%%%%%%%%%%%%%%%%%%%%%%%

In the presence of 2HDMs with MFV, the observables in $B \to K^* \gamma$ decays are also expected to be correlated with each other. Within the parameter space allowed by $\mathcal B(B \to X_s \gamma)$, these are shown in figure~\ref{fig:corr:B2KVg} for both the type-III and the type-C model. The correlations between the two branching ratios are trivial and similar between the two models; the large values for the branching ratios correspond to the case where the NP contribution to $C_{7}^{\mathrm{eff}}$ is about twice the size of the SM one. For the direct CP asymmetries, the allowed ranges in the type-C model are much larger than the ones in the type-III model, which is similar to the case observed in the inclusive decays. For the isospin asymmetry, on the other hand, the allowed ranges in the type-C model are much smaller than the ones in the type-III model. Very large values for $\Delta(K^*\gamma)$ correspond to the case where a strong cancellation between the SM and the NP contributions to $C_{7}^{\mathrm{eff}}(\mu_b)$ occurs, making the remaining parts, such as the annihilation and spectator-scattering contributions, relatively important.

Thus, it is concluded that the direct CP and isospin asymmetries in $B\to K^*\gamma$ decays could provide constraints on the parameter space in a way complementary to the branching ratios. Improved experimental measurements and theoretical predictions will make these observables more powerful for exploring NP.

\subsection{$B \to \rho \gamma$ decays in 2HDM with MFV}

For the exclusive $B\to \rho \gamma$ decays, since the CKM factors $\lambda_u^{(d)}$ and $\lambda_t^{(d)}$ are comparable in magnitude, the two decay amplitudes ${\cal C}_7^{(u)}$ and ${\cal C}_7^{(t)}$ should be included simultaneously. This feature makes these decays particularly interesting in constraining the CKM unitarity triangle~\cite{PQCD:2,PQCD:9,QCDF:2,QCDF:4,QCDF:7} and probing physics beyond the SM~\cite{Lyon:2013gba,Lee:2006qv,Kim:2004zm,Jung:2012vu,Li}.

%%%%%%%%%%%%%%%%%%%%%%%%%%%%%%%%%%%%%%%%%%%%%%%%%%%%%%%%%%%%%%%%%%%
\begin{figure}[t]
  \centering
  \subfigure[real couplings]{\includegraphics[width=0.91\textwidth]{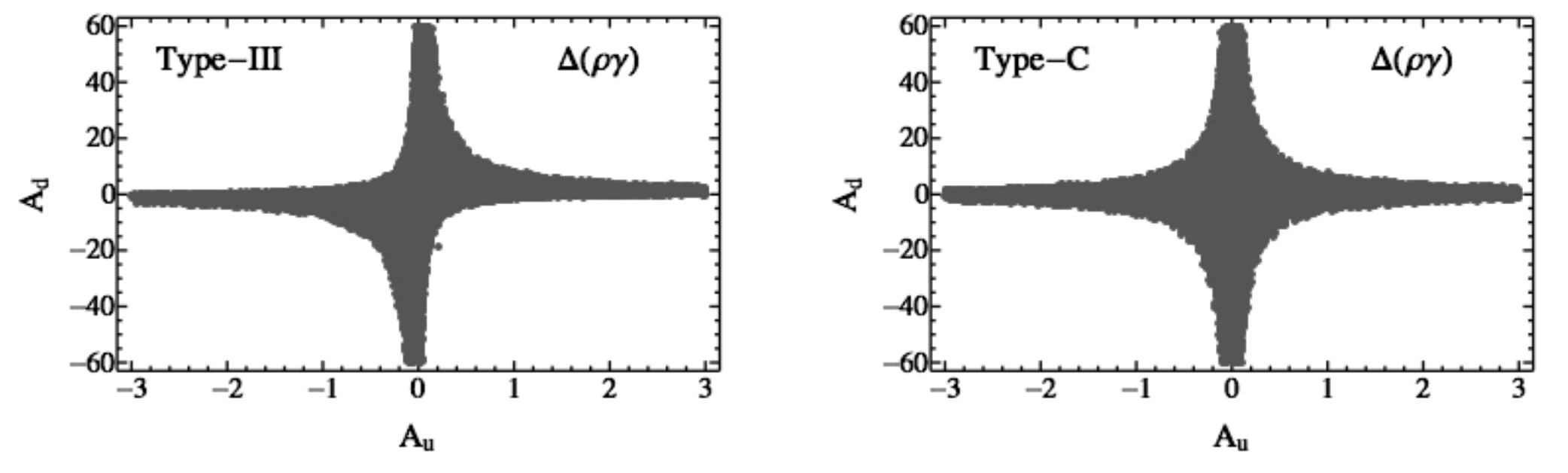}}
  \subfigure[complex couplings]{\includegraphics[width=0.92\textwidth]{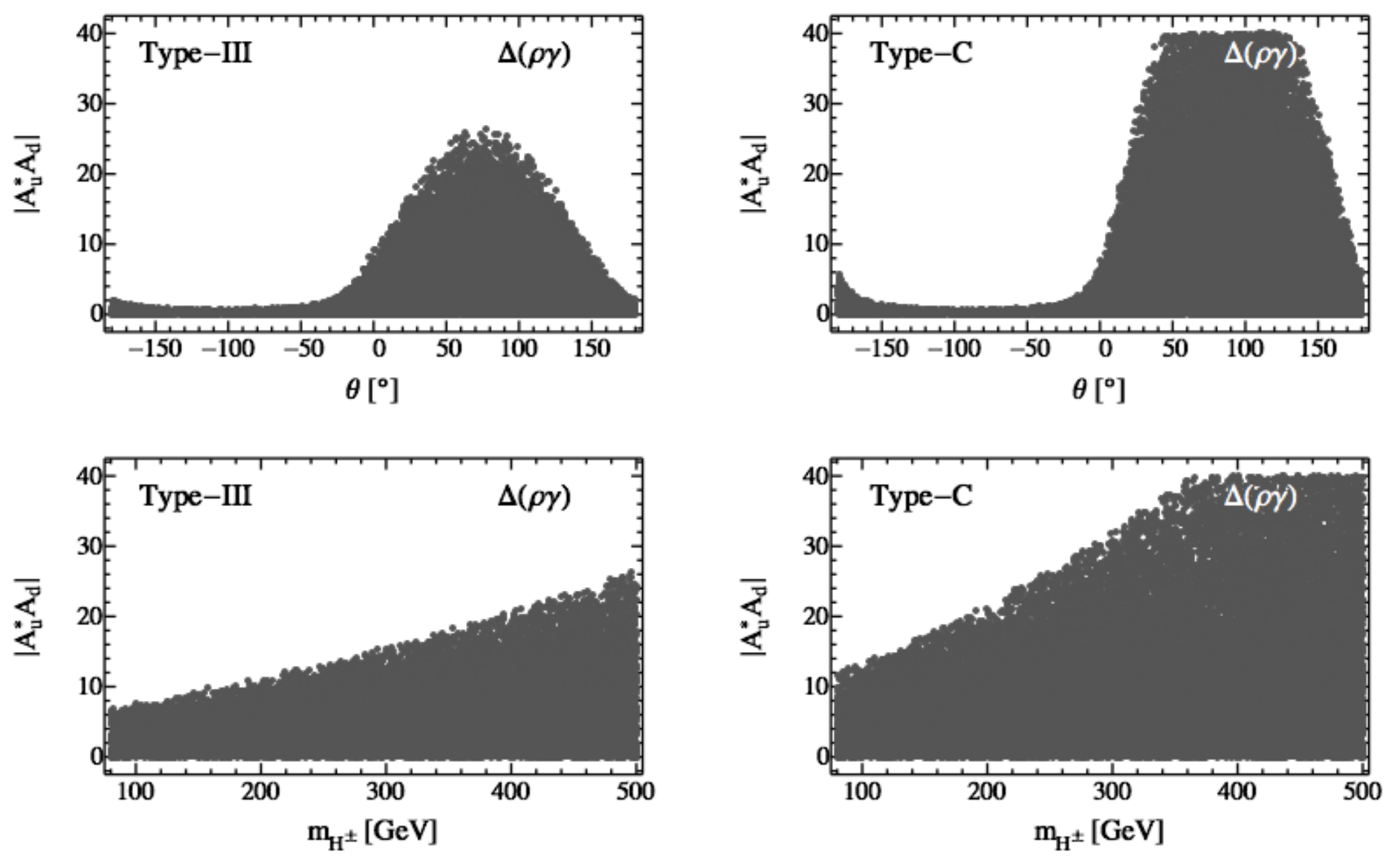}}
  \caption{\small Constraints on the couplings $A_u$ and $A_d$ from $\Delta(\rho\gamma)$. The other captions are the same as in figure~\ref{fig:B2Xsg}.}
  \label{fig:Rho}
\end{figure}
%%%%%%%%%%%%%%%%%%%%%%%%%%%%%%%%%%%%%%%%%%%%%%%%%%%%%%%%%%%%%%%%%%%

%%%%%%%%%%%%%%%%%%%%%%%%%%%%%%%%%%%%%%%%%%%%%%%%%%%%%%%%%%%%%%%%%%%
\begin{figure}[htbp]
  \centering
  \includegraphics[width=0.98\textwidth]{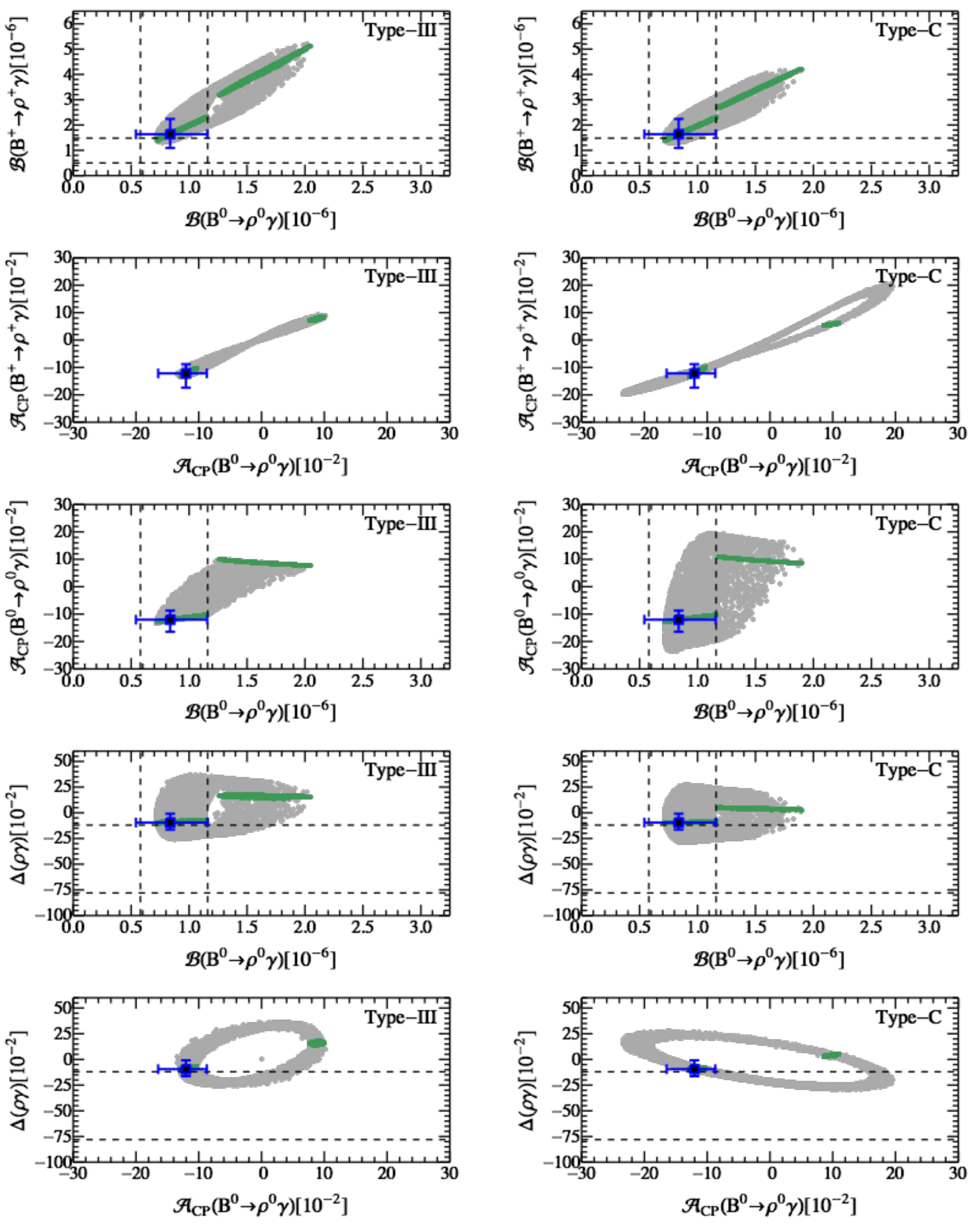}
  \caption{\small Correlation plots between the observables of $B \to \rho \gamma$ decays within the parameter space allowed by $\mathcal B(B \to X_s \gamma)$. The other captions are the same as in figure~\ref{fig:corr:B2Xg}.}
  \label{fig:corr:B2Rhog}
\end{figure}
%%%%%%%%%%%%%%%%%%%%%%%%%%%%%%%%%%%%%%%%%%%%%%%%%%%%%%%%%%%%%%%%%%%

Specific to the type-III and type-C models, it is found that, similarly to the case of $B\to K^*\gamma$ decays, constraints from the two branching ratios are also slightly weaker than the ones from $\mathcal B(B \to X_s \gamma)$, and the two direct CP asymmetries are again found not to be able to put any constraints on the model parameters. Thus, we only show in figure~\ref{fig:Rho} the constraints on the Yukawa couplings $A_u$ and $A_d$ from the isospin asymmetry $\Delta(\rho\gamma)$. From these plots, the following observations are made:
\begin{itemize}
      \item For both the type-III and the type-C model, the isospin asymmetry $\Delta(\rho\gamma)$ exhibits a different dependence on the phase $\theta$ from the branching ratios, but also the observable $\Delta(K^*\gamma)$. This is due to the comparable contribution from the extra term proportional to $\lambda_u^{(d)}$, which is associated with an extra weak phase $\arg(V_{ub})$.

      \item For the type-C model, unlike the case in $B \to K^* \gamma$ decays, the isospin asymmetry $\Delta(\rho\gamma)$ excludes most of the regions with large values of $|A_u|$ and $|A_d|$. This is mainly due to the discrepancy between the experimental measurement and the SM prediction, in which the current data is quite below the SM prediction. In the region with large $|A_u|$ and $|A_d|$, although still being opposite in sign, the obtained values of $\Delta(\rho\gamma)$ in both cases are larger than the experimental data, and hence the corresponding regions are excluded.

      \item Due to their different dependence on the phase $\theta$, the combined constraints from the branching ratios and the two isospin asymmetries should be more stringent, which will be explored in detail in section~\ref{sec:concl}.
\end{itemize}

Within the parameter space allowed by $\mathcal B(B \to X_s \gamma)$, correlations between the observables of $B \to \rho \gamma$ decays are shown in figure~\ref{fig:corr:B2Rhog}. It is observed that, for the direct CP asymmetries, the predicted ranges relative to the SM predictions are different between the $B\to K^* \gamma$ and $B \to \rho \gamma$ decays. This difference is similar to that observed in the inclusive $B\to X_{s,d}\gamma$ decays and the reason is almost the same.

%%%%%%%%%%%%%%%%%%%%%%%%%%%%%%%%%%%%%%%%%%%%%%%%%%%%%%%%%%%%%%%%%%%
\begin{figure}[t]
  \centering
  \includegraphics[width=0.98\textwidth]{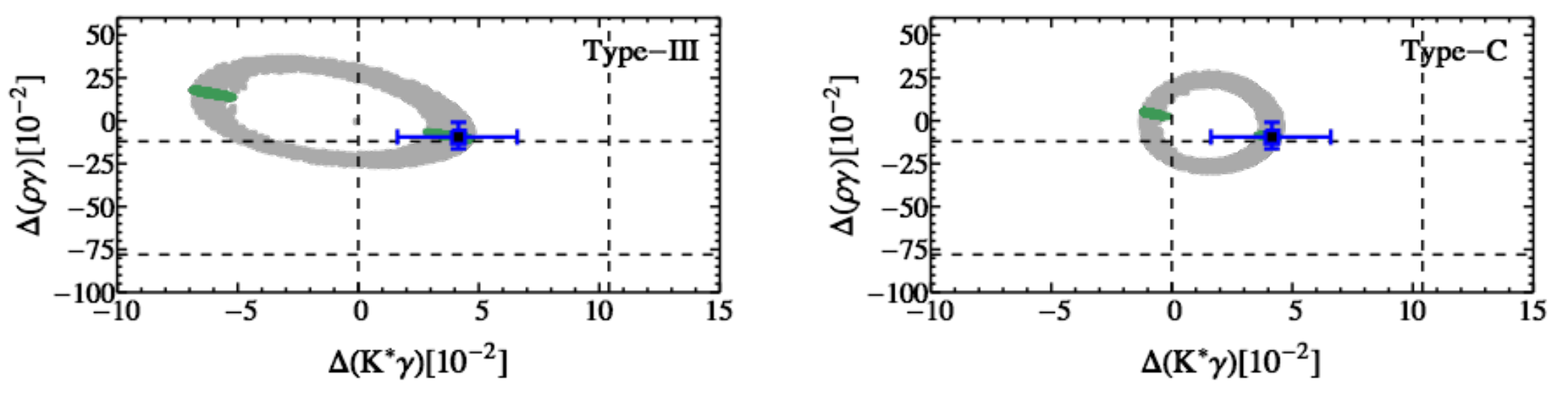}
  \caption{\small Correlation plots between the two isospin asymmetries $\Delta(K^*\gamma)$ and $\Delta(\rho\gamma)$ within the parameter space allowed by $\mathcal B(B \to X_s \gamma)$. The other captions are the same as in figure~\ref{fig:corr:B2Xg}.}
  \label{fig:corr:Isospin}
\end{figure}
%%%%%%%%%%%%%%%%%%%%%%%%%%%%%%%%%%%%%%%%%%%%%%%%%%%%%%%%%%%%%%%%%%%

We also show in figure~\ref{fig:corr:Isospin} the correlation between the two isospin asymmetries $\Delta(K^*\gamma)$ and $\Delta(\rho\gamma)$. It can be seen that, even under the constraint from $\mathcal B(B \to X_s \gamma)$, large values for the two isospin asymmetries remain allowed relative to their respective SM predictions, with much wider ranges for $\Delta(K^*\gamma)$ than for $\Delta(\rho\gamma)$. To understand this, we should note that the allowed ranges of the isospin asymmetry under the 2HDMs are determined by the NP effects with small values of $|A_u^*A_d|$. In this region, the contribution from the weak annihilation and the spectator scattering through the current-current operator $\mathcal O_2$ plays an important role, which corresponds to the term proportional to ${\rm Re}(\lambda_u^{(D)}/\lambda_t^{(D)}\cdot C_2/C_7^{\rm eff})$. Within the SM, this term is doubly Cabibbo suppressed for the $B \to K^* \gamma$, but no suppression for the $B \to \rho \gamma$ decays. Thus, due to the absence of the CKM suppression, much wider ranges for $\Delta(K^*\gamma)$ are predicted than for $\Delta(\rho\gamma)$ in the two NP models. This also explains why the predicted ranges for the direct CP and isospin asymmetries do not show a large difference between the two models, complemented by the fact that the two models give similar corrections to $C_7^{\rm eff}(\mu_b)$. However, with the constraint from $\mathcal B(B \to X_s \gamma)$ taken into account, the very large central value of $\Delta(\rho\gamma)$ cannot be accommodated by the two models.

From the above observations, we conclude that, similarly to $B\to K^*\gamma$ decays, the isospin asymmetry $\Delta(\rho\gamma)$ is also a very important observable in constraining the charged-Higgs Yukawa couplings. A confirmation of the present central value with higher precision would challenge the SM as well as the two models considered here.

\subsection{Other $B \to V \gamma$ decays in 2HDM with MFV}

In this subsection, we discuss the other exclusive radiative B-meson decays, including $B^0 \to \omega \gamma$ and $B_s \to \bar K^{*0} \gamma$ induced by $ b \to d$ transition, as well as $B_s \to \phi \gamma$ induced by $b \to s$ transition. At present, only the branching ratios of $B_s \to \phi \gamma$ and $B^0 \to \omega \gamma$ have been measured, which however could not provide stronger constraints on the model parameters than the ones from $\mathcal B(B \to X_s \gamma)$. Here we do not consider the pure annihilation-dominated $B\to V\gamma$ decays, which are predicted to be very tiny within the QCDF formalism, $\sim \mathcal{O}(10^{-10})$~\cite{Li:2003kz,Hua:2010we}. Furthermore, a quantitative discussion does not seem appropriate in this case, as the two models considered here do not imply large enhancements.

%%%%%%%%%%%%%%%%%%%%%%%%%%%%%%%%%%%%%%%%%%%%%%%%%%%%%%%%%%%%%%%%%%%
\begin{figure}[htbp]
  \centering
  \includegraphics[width=0.98\textwidth]{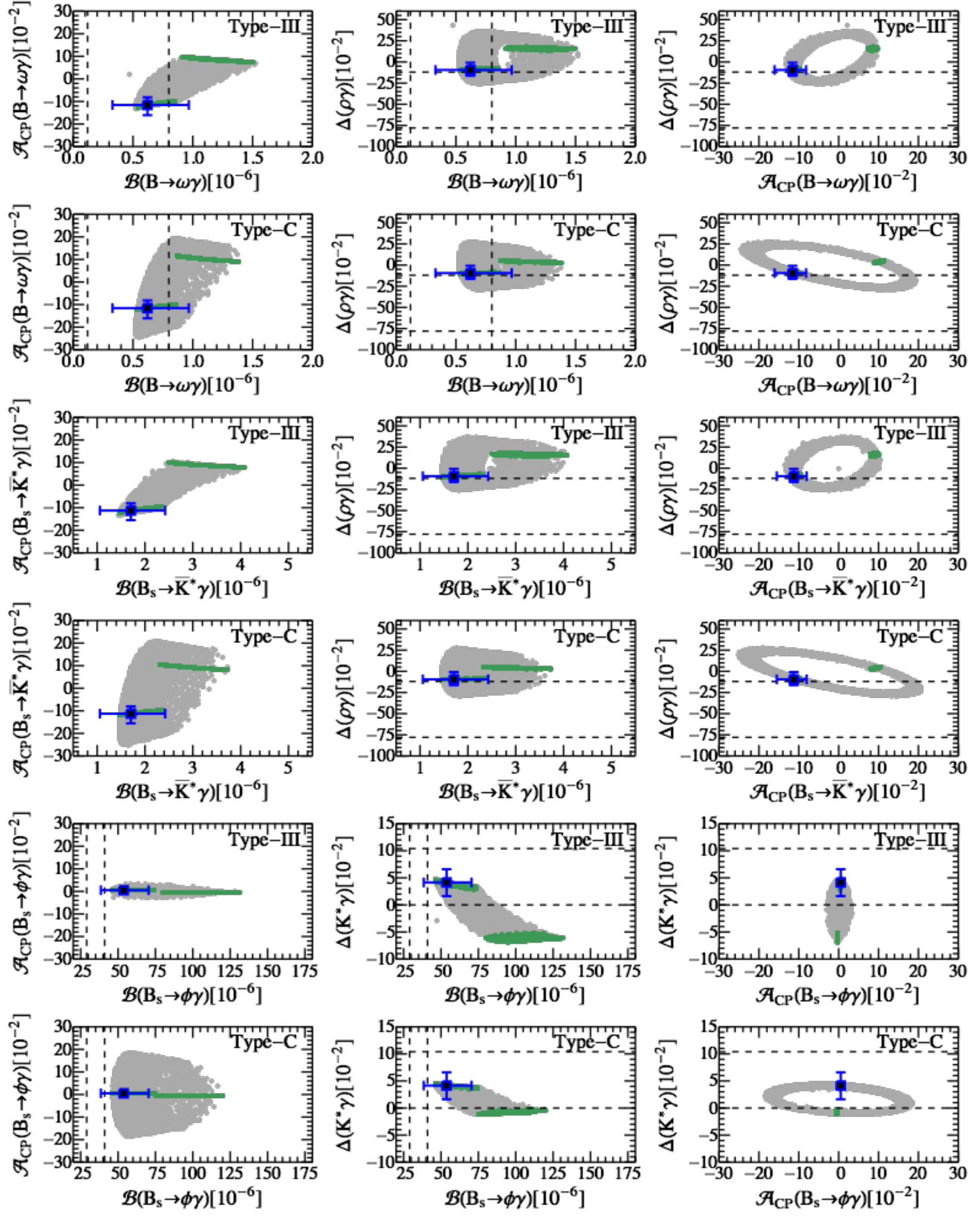}
  \caption{\small Correlation plots between the observables of $B \to \omega \gamma$, $B_s \to \bar K^{*0}\gamma$ and $B \to \phi \gamma$ decays, as well as $\Delta(K^*\gamma)$ and $\Delta(\rho\gamma)$. The other captions are the same as in figure~\ref{fig:corr:B2Xg}.}
  \label{fig:corr:B2Vg}
\end{figure}
%%%%%%%%%%%%%%%%%%%%%%%%%%%%%%%%%%%%%%%%%%%%%%%%%%%%%%%%%%%%%%%%%%%

Since neither the direct CP asymmetries of these decays nor the observables of $B_s\to \bar{K}^*\gamma$ have been measured so far, we show in figure~\ref{fig:corr:B2Vg} the predicted correlations among the various observables in these $B \to V \gamma$ decays as well as the two isospin asymmetries $\Delta(K^*\gamma)$ and $\Delta(\rho\gamma)$, within the parameter space allowed by the observable $\mathcal B(B \to X_s \gamma)$. With improved measurements from the LHCb and the future Super-B factory, these interplays may provide useful information about the NP model.

It is interesting to note that, even after taking into account the constraints obtained from the previous sections, large derivations from the SM predictions for the direct CP asymmetries, especially in the case of type-C model, are still possible. As pointed in refs.~\cite{Jung:2012vu,PQCD:10}, the direct CP asymmetry of $B_s \to \phi \gamma$ decay is predicted to be relatively tiny within the SM and does not suffer large hadronic uncertainties, which makes it particularly sensitive not only to the 2HDMs considered here, but also to every model introducing new weak phases in $b\to s$ transitions.

\section{Conclusions}
\label{sec:concl}

In the ``Higgs basis" for a generic 2HDM, only one doublet gets a nonzero vacuum expectation value and, under the MFV criterion, the other one is fixed to be either colour-singlet or colour-octet, referred to, respectively, as the type-III and type-C models. Due to the absence of FCNC transitions, both of these two models imply very interesting phenomena in some low-energy processes. In this paper, we have studied their effects on the exclusive radiative B-meson decays due to the exchange of colourless or coloured charged-Higgs boson. Our main conclusions can be summarized as follows:
\begin{itemize}
  \item Constraints from the branching ratios of exclusive decays are slightly weaker than the ones from the inclusive $B \to X_s \gamma$ decay. As the branching ratio is proportional to $|C_7^{\rm eff}(\mu_b)|^2$ to the first order, for which the NP contributions make no significant differences between the two models, constraints from these observables are almost indistinguishable for the type-III and the type-C model.

  \item Complementary constraints on the model parameters can be obtained from the two isospin asymmetries $\Delta(K^*\gamma)$ and $\Delta(\rho\gamma)$, which vary like $1/C_{7}^{\mathrm{eff}}(\mu_b)$ at the LO. Especially, once constraints from the current data on $\Delta(K^*\gamma)$ are taken into account, the allowed regions in which the NP contribution is about twice the size of the SM one are already excluded.

  \item As the two models predict similar magnitudes but with opposite signs for $C_8^{\rm eff}(\mu_b)$, the direct CP and isospin asymmetries of $b\to s$ processes, to both of which the main SM contributions are doubly Cabibbo suppressed, are found to be more suitable for discriminating the two models. Due to the absence of CKM suppression, on the other hand, contributions from the term proportional to $\lambda_u^{(d)}$ makes these observables of $b\to d$ processes less sensitive to $C_8^{\rm eff}(\mu_b)$ and exhibit a different dependence on the phase $\theta$.

  \item Since most of the observables still suffer large uncertainties, we have also investigated correlations between the observables in exclusive $B\to V\gamma$ and inclusive $B\to X_{s,d}\gamma$ decays, within the parameter space allowed by $\mathcal B(B \to X_s \gamma)$. Some of them will become relevant with the advent of more precise experimental data and theoretical predictions.
\end{itemize}

\begin{figure}[t]
  \centering
  \subfigure[real couplings]{\includegraphics[width=0.91\textwidth]{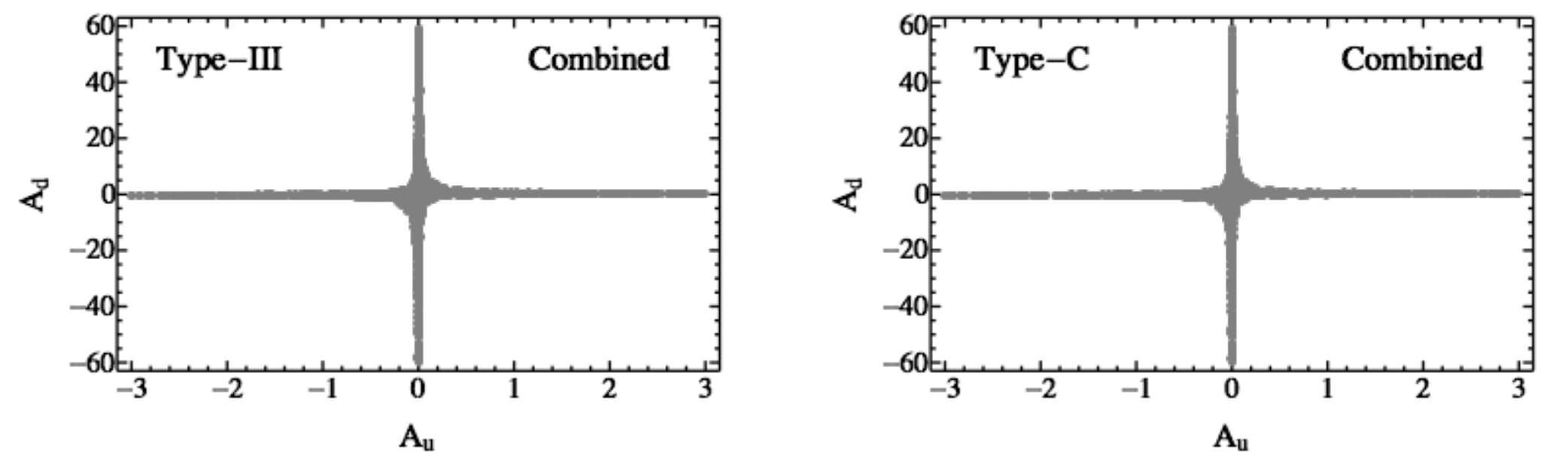}}
  \subfigure[complex couplings]{\includegraphics[width=0.92\textwidth]{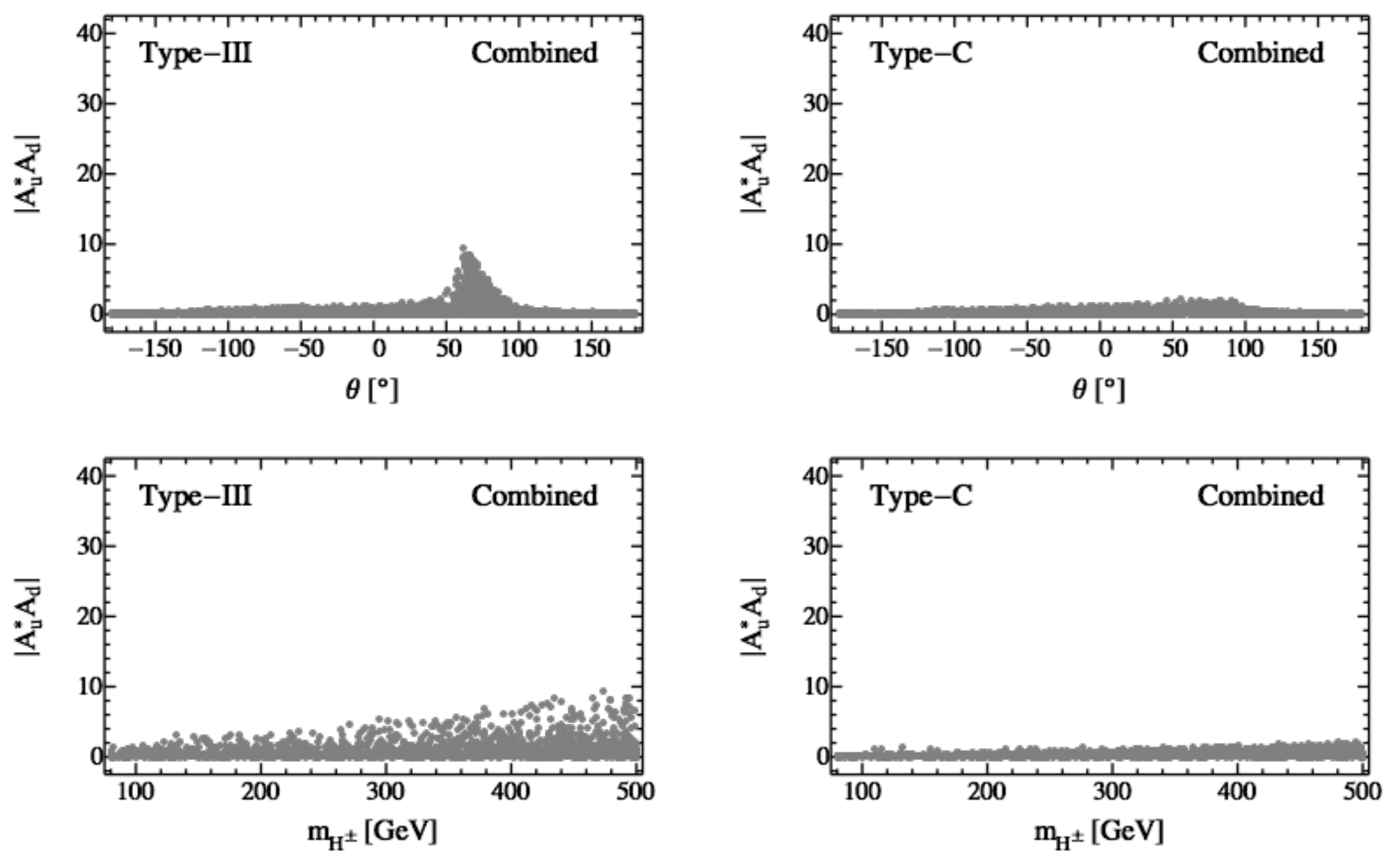}}
  \caption{\small Combined constraints on the couplings $A_u$ and $A_d$ from all the available experimental data on $B \to V \gamma$ and $B \to X_{s,d} \gamma$ decays. The other captions are the same as in figure~\ref{fig:B2Xsg}.}
  \label{fig:combined}
\end{figure}

To see clearly the complementary effects between these observables, we show in figure~\ref{fig:combined} the final combined constraints from all the available experimental data on $B \to V \gamma$ and $B \to X_{s,d} \gamma$ decays. With respect to the regions shown in figure~\ref{fig:B2Xsg}, one can see that constraints from the exclusive observables could exclude a significant additional part of the parameter space.

%%%%%%%%%%%%%%%%%%%%%%%%%%%%%%%%%%%%%%%%%%%%%%%%%%%%%%%%%%%%%%%%%%%
%%%%%%%%%%%%%%%%%%%%%%%%%%%%%%%%%%%%%%%%%%%%%%%%%%%%%%%%%%%%%%%%%%%

With the experimental progress expected from the LHCb and the future Super-B factory, as well as the improved theoretical predictions for these decays, either constraints shown here will be strengthened or signs of non-standard effects rather than the ones considered here will show up. Of special interest in this respect are the two isospin asymmetries.

\section*{Acknowledgements}

The work was supported by the National Natural Science Foundation of China~(NSFC) under contract Nos.~11005032, 11075059, 11225523 and 11221504. X.~Q. Li was also supported by the Specialized Research Fund for the Doctoral Program of Higher Education of China~(Grant Nos.~20104104120001) and by the Scientific Research Foundation for the Returned Overseas Chinese Scholars, State Education Ministry. X.~B. Yuan was also supported by the CCNU-QLPL Innovation Fund~(QLPL2011P01) and the Excellent Doctorial Dissertation Cultivation Grant from Central China Normal University.

\end{document}